\documentclass{article}

\usepackage[utf8]{inputenc}
\DeclareUnicodeCharacter{03B5}{$\epsilon$}

\usepackage[left=2cm,
            right=2cm,
            top=2.25cm,
            bottom=2.25cm,
            headheight=12pt,
            letterpaper]{geometry}
\usepackage{calc}

\usepackage{times}      
\usepackage{mathptmx}   
\usepackage{microtype}
\usepackage{setspace}   

\usepackage{amsmath,amsfonts,amssymb}
\usepackage{siunitx}    

\usepackage[dvipsnames]{xcolor}
\usepackage{graphicx}
\usepackage[inkscapeformat=png]{svg} 
\usepackage{adjustbox}

\usepackage{booktabs}
\usepackage{float}
\usepackage{newfloat}

\usepackage{enumitem}

\usepackage{ifpdf}
\usepackage{ifthen}

\usepackage{authblk}

\usepackage[explicit]{titlesec}

\usepackage{fancyhdr} 
\usepackage{lastpage} 

\usepackage{xurl}

\usepackage[colorlinks=true, allcolors=blue]{hyperref}

\usepackage{newfloat}
\DeclareFloatingEnvironment[]{suppfigure}
\DeclareFloatingEnvironment[]{supptable}
\DeclareFloatingEnvironment[]{extendedfigure}
\DeclareFloatingEnvironment[]{tempfigure}

\usepackage[labelfont={bf,sf},
            labelsep=period,
            justification=raggedright]{caption}

\DeclareCaptionLabelSeparator{pipe}{ $\vert$ }
\usepackage[defernumbers=true, style=nature]{biblatex}

\usepackage{algorithm}
\usepackage{algpseudocode}

\usepackage[most]{tcolorbox}

\usepackage{minted}

\usepackage{enumitem}

\newcommand{\keywords}[1]{\def\@keywords{#1}}

\makeatletter
\def\xabstract{abstract}
\long\def\abstract#1\end#2{\def\two{#2}\ifx\two\xabstract 
\long\gdef\theabstract{\ignorespaces#1}
\def\go{\end{abstract}}\else
\typeout{^^J^^J PLEASE DO NOT USE ANY \string\begin\space \string\end^^J
COMMANDS WITHIN ABSTRACT^^J^^J}#1\end{#2}
\gdef\theabstract{\vskip12pt BADLY FORMED ABSTRACT: PLEASE DO
NOT USE {\tt\string\begin...\string\end} COMMANDS WITHIN
THE ABSTRACT\vskip12pt}\let\go\relax\fi
\go}
\renewcommand{\@maketitle}{%
{%
\thispagestyle{empty}%
\vskip-36pt%
{\raggedright\sffamily\bfseries\fontsize{20}{25}\selectfont \@title\par}%
\vskip10pt
{\raggedright\sffamily\fontsize{12}{16}\selectfont  \@author\par}
\vskip18pt%
{%
\noindent
{\parbox{\dimexpr\linewidth-2\fboxsep\relax}{\color{color1}\large\sffamily\textbf{ABSTRACT}}}
}%
\vskip10pt
{%
\noindent
\colorbox{color2}{%
\parbox{\dimexpr\linewidth-2\fboxsep\relax}{%
\sffamily\small\textbf\\\theabstract
}%
}%
}%
\vskip25pt%
}%
}
\makeatother
\definecolor{color1}{RGB}{0,0,0} 
\definecolor{color2}{gray}{1} 
\titleformat{\section}
    {\color{color1}\large\sffamily\bfseries}
    {\thesection}
    {0.5em}
    {#1}
    []
\titleformat{name=\section,numberless}
    {\color{color1}\large\sffamily\bfseries}
    {}
    {0em}
    {#1}
    []    
\titleformat{\subsection}
    {\sffamily\bfseries}
    {\thesubsection}
    {0.5em}
    {#1}
    []
\titleformat{\subsubsection}
    {\sffamily\small\bfseries\itshape}
    {\thesubsubsection}
    {0.5em}
    {#1}
    []        
\titleformat{\paragraph}[runin]
    {\sffamily\small\bfseries}
    {}
    {0em}
    {#1}  
\titlespacing*{\section}{0pc}{3ex plus 4pt minus 3pt}{5pt}
\titlespacing*{\subsection}{0pc}{2.5ex plus 3pt minus 2pt}{0pt}
\titlespacing*{\subsubsection}{0pc}{2ex plus 2.5pt minus 1.5pt}{0pt}
\titlespacing*{\paragraph}{0pc}{1.5ex plus 2pt minus 1pt}{10pt}

\makeatletter
\newcommand\setcurrentname[1]{\def\@currentlabelname{#1}}
\makeatother

\newif\ifprintcomments

\printcommentsfalse

\newif\ifprinthighlight

\newcommand{\rebuttal}[1]{
	\ifprinthighlight
		{\textcolor{OliveGreen}{#1}}
	\else
		{#1}
	\fi
}

\newcommand{\rebuttaltemp}[1]{
	\ifprinthighlight
		{\textcolor{red}{#1}}
	\else
	\fi
}

\printhighlightfalse

\title{Explainable AI for computational pathology identifies model limitations and tissue biomarkers}

\author[1,2,3,+]{Jakub R. Kaczmarzyk}
\author[4,+]{Chanwoo Kim}
\author[4,+]{Soham Gadgil}
\author[5]{Deepika Savant}
\author[5,6]{Zhen Zhao}
\author[1]{Joel H. Saltz}
\author[4,*]{Su-In Lee}
\author[2,*]{Peter K. Koo}
\affil[1]{Department of Biomedical Informatics, Stony Brook University, Stony Brook, NY, USA}
\affil[2]{Simons Center for Quantitative Biology, Cold Spring Harbor Laboratory, Cold Spring Harbor, NY, USA}
\affil[3]{Medical Scientist Training Program, Stony Brook University, Stony Brook, NY, USA}
\affil[4]{Paul G. Allen School of Computer Science and Engineering, University of Washington, Seattle, WA, USA}
\affil[5]{Northwell Health, Greenvale, NY, USA}
\affil[6]{Cold Spring Harbor Laboratory, Cold Spring Harbor, NY, USA}
\affil[+]{These authors contributed equally to this work.}
\affil[*]{Corresponding authors: koo@cshl.edu and suinlee@cs.washington.edu}

\date{}

\begin{document}

\begin{refsegment}

\begin{abstract}
Deep learning models show promise in digital pathology, but their opaque decision-making processes undermine trust and limit their widespread clinical adoption. \rebuttal{To address this challenge, we present HIPPO, an explainable AI method for analyzing weakly-supervised multiple-instance learning (MIL) models that are widely used in whole slide image analysis. HIPPO constructs counterfactual whole slide images by systematically removing or adding selected tissue regions, providing a principled way to quantify how specific histologic areas influence model predictions under the MIL framework. This capability enables rigorous model interpretation, quantitative hypothesis testing, bias detection, and performance evaluation that extend beyond standard metrics.} We demonstrate HIPPO’s capabilities across numerous models performing clinically important tasks, such as breast metastasis detection in axillary lymph nodes, prognostication in breast cancer and melanoma, and \textit{IDH} mutation classification in gliomas. In metastasis detection, HIPPO uncovered critical model limitations that were undetectable by standard performance metrics or attention-based methods. For prognostic prediction, HIPPO outperformed attention by providing more nuanced insights into tissue elements influencing outcomes. In a proof-of-concept study, HIPPO facilitated hypothesis generation for identifying melanoma patients who may benefit from immunotherapy. In \textit{IDH} mutation classification, HIPPO more robustly identified the pathology regions responsible for false negatives compared to attention, suggesting its potential to outperform attention in explaining model decisions. In summary, HIPPO expands the explainable AI toolkit for computational pathology by enabling deeper insights into model behavior. This framework supports the trustworthy development, deployment, and regulation of weakly-supervised models in clinical and research settings, promoting their broader adoption in digital pathology.
\end{abstract}

\maketitle

\section*{Introduction}

Digital pathology has emerged as a transformative force in medicine, ushering in an era where computational methods can augment and enhance the diagnostic and prognostic capabilities of pathologists. By digitizing whole slide images (WSIs) of tissue specimens, this field has opened new avenues for applying advanced machine learning techniques to analyze complex histological patterns and features. The potential impact of computational pathology is far-reaching, promising to improve diagnostic accuracy, standardize interpretation, and uncover novel biomarkers that may inform personalized treatment strategies \autocite{van2021deep, echle2021deep, topol2019high, song2023artificialcpath, niazi2019digital, rakha2021current, cui2021artificial, morales2021artificial, tizhoosh2018artificial, tran2021deep, cifci2023ai, shmatko2022artificial, perez2024guide, unger2024systematic, bera2019artificial}.

\begin{figure}[H]
    \vspace{-1.2cm}
    \centering
    \includegraphics[width=6.5in, clip, trim=0cm 3.5cm 0cm 0cm]{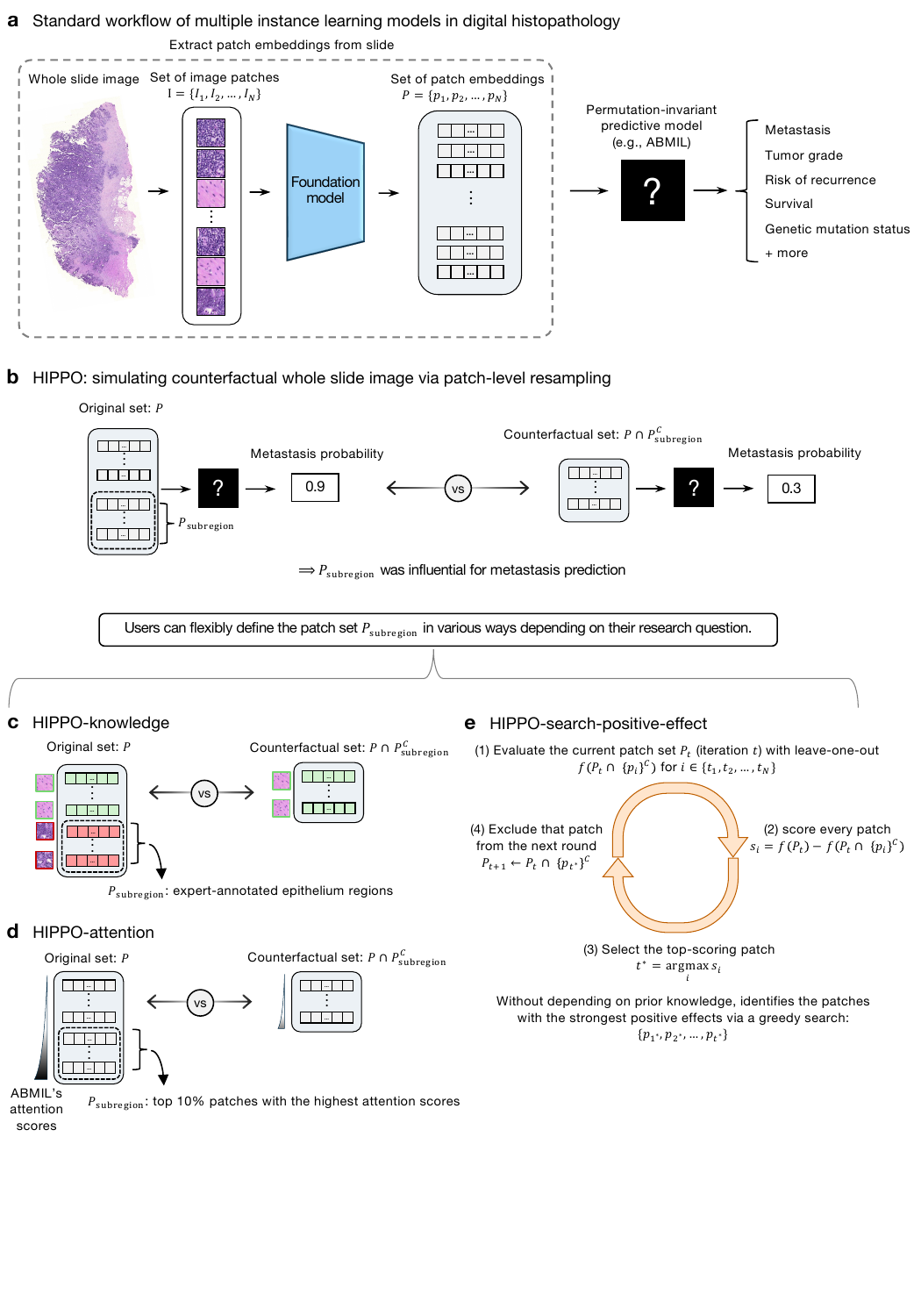}
    \rebuttaltemp{Modified figure: Clarified the workings of HIPPO and its variants using illustrative diagrams}
    \caption{
    \small{
    \rebuttal{\textbf{Overview of HIPPO explainable AI toolkit.} HIPPO enables quantitative assessment of how specific tissue regions impact model predictions, enhancing interpretation and validation of AI models. 
    \textbf{a,} Standard workflow of multiple instance learning (MIL) models. A whole slide image (WSI) is divided into patches and embedded using a pretrained foundation model. A task-specific MIL model then aggregates the patch embeddings to predict a specimen-level label. These MIL models are permutation-invariant, treating patches as an unordered set.
    \textbf{b,} HIPPO simulates counterfactual WSIs through patch-level resampling, thereby revealing how a subregion in WSI influences the model's prediction. HIPPO constructs a counterfactual by ablating patches from the original set, evaluates the model on this counterfactual set, and compares the prediction to that of the original WSI to quantify the effect of the subregion. Users can flexibly define a subregion to ablate $P_\text{subregion}$ depending on their research question.
    \textbf{c,} \textit{HIPPO-knowledge}: choose patches to ablate or add based on expert annotations to test if the patches that drive the model's prediction align with expert knowledge.
    \textbf{d,} \textit{HIPPO-attention}: patches with the highest attention scores assigned by the MIL model are removed to evaluate whether high-attention patches truly drive the model's prediction.
    \textbf{e,} \textit{HIPPO-search-positive-effect}: a greedy leave-one-out procedure iteratively scores and removes patches to automatically discover the set of patches that most increases the model's predictive output, without depending on prior knowledge.}
    }}
\label{fig:hippo-schematic}
\end{figure}

Building on these advances, the development of pathology foundation models represents a leap forward in computational pathology. These models provide powerful, generalizable tools for analyzing histological features \autocite{azizi2022remedis, wang2022ctranspath, filiot2023phikon, wang2023retccl, chen2024uni, lu2024visualconch, xu2024whole, wang2024pathology, vorontsov2024virchow, zimmermann2024virchow2, filiot2024phikonv2, nechaev2024hibou, aben2024kaiko, vorontsov2024virchow, zimmermann2024virchow2}. Trained on vast collections of histology image patches in a self-supervised manner, they learn general-purpose representations of histological features. By focusing on patches, foundation models can efficiently process the large size of WSIs while capturing key local patterns. The resulting patch-level representations can then be used as inputs to downstream task-specific models. For specimen-level prediction tasks, these embeddings are often integrated into a multiple-instance learning (MIL) framework, such as attention-based MIL (ABMIL) architecture \autocite{ilse2018abmil}. In this setup, the foundation model provides patch-level embeddings, and ABMIL \rebuttal{identifies relevant patches and assigns attention weights to aggregate them, finally obtaining specimen-level predictions (Fig. \ref{fig:hippo-schematic}a).} This combination has quickly become the standard for leveraging foundation models in specimen-level classification, offering a scalable and effective way to utilize their pre-trained knowledge for complex clinical and research applications. Unlike traditional image classifiers, MIL models only require labels for a bag of image patches (i.e., specimen-level labels) for training, enabling scalable application across diverse tasks \autocite{gadermayr2024multiple}. This flexibility has enabled MIL-based workflows to excel across a range of tasks, including cancer detection \autocite{bejnordi2017camelyon16, bulten2022panda}, diagnosis \autocite{qiu2021attention, butke2021end, del2022constrained, del2021attention}, identification of primary cancer origin \autocite{lu2021ai}, grading \autocite{bulten2022panda, su2022attention2majority, li2021prostategrading}, genomic aberration detection \autocite{saldanha2023self, dernbach2024dissecting, zheng2024predicting, schirris2021deepsmile}, molecular phenotyping \autocite{valieris2024weakly, boehm2024oncotype, el2024regression}, treatment response prediction \autocite{lipkova2022crane, jiang2023biology, mallya2024benchmarking}, and prognostication \autocite{chen2021multimodal, chen2022porpoise, jiang2023biology, ammeling2023attention}. 

However, the widespread adoption of ABMIL models in clinical settings is hindered by challenges in model interpretability and trustworthiness \autocite{plass2023explainability, castro2020causality, tizhoosh2018artificial, tran2021deep, foote2021now, ghaffari2022adversarial, dawood2024buyer}. General-purpose explainable AI methods, such as LIME \autocite{ribeiro2016lime}, SHAP \autocite{lundberg2017shap}, and SmoothGrad \autocite{selvaraju2017grad}, have been proposed to address these challenges by providing insights into model behavior and identifying features that may influence predictions. These methods often rely on linear or additive approximations, which can oversimplify complex model behaviors \autocite{lipton2018mythos, bilodeau2024impossibility}. In ABMIL models, attention scores, which are used as weights for aggregating patch embeddings, have emerged as a widely used approach for interpreting ABMIL models, as evidenced by their extensive use in recent studies \autocite{ilse2018abmil, lipkova2022crane, lu2021ai, chen2022porpoise, lu2021clam, xiong2023diagnose, wang2024pathology, niehues2023generalizable, el2024regression, graziani2022attention, cai2024pathologist, yao2024application, zhang2023attention, sehring2023leveraging}. Although attention highlights regions of interest within WSIs, it does not directly quantify the influence of these regions on model predictions \autocite{javed2022additivemil, nan2024establishing, jain2019attention}. This gap between attention and predictive influence can lead to misinterpretation, undermining trust in model decisions and limiting clinical applicability \autocite{liu2024attention, raff2024reproducibility, lin2023interventional, pocevivciute2020survey}.  There is a pressing need for methods that explicitly evaluate how specific tissue regions contribute to model outputs.

\rebuttal{One principled approach is to create counterfactual WSIs in which the subregion containing the feature of interest is ablated. Such counterfactual inputs would allow us to explore ``what if'' scenarios by observing the differential impact on model behavior. For example, simulating WSIs with larger or smaller tumor regions could help determine the model's sensitivity to tumor size, while modifying regions with tumor-infiltrating lymphocytes might reveal their role in influencing predictions, such as for prognosis or treatment response. However, obtaining reliable counterfactual WSIs presents significant technical challenges. Occlusion-based methods (e.g., masking regions with gray boxes \autocite{zeiler2014visualizing}) can produce unnatural artifacts and cause erratic behavior in the model, as these patterns were not encountered during training, leading to interpretations that may be difficult to trust. To remove specific information from an image while keeping it realistic, generative models offer a potential solution \autocite{pozzi2024synthetic}, but applying robust inpainting or image synthesis at gigapixel scales for WSIs to fill in the modified regions realistically remains a technical challenge. The ability to generate realistic and valid counterfactual WSIs for clinical use has yet to be proven.} Consequently, practical and trustworthy counterfactual explanations for computational histopathology models remain a major unmet need.

To address these challenges, we introduce HIPPO (Histopathology Interventions of Patches for Predictive Outcomes), an explainable AI method designed to enhance trust in pathology models and provide deeper insights into their decision-making processes (\nameref{sec:methods}). \rebuttal{HIPPO simulates counterfactual WSIs by modifying the set of patches that compose a slide, selectively removing regions or adding new ones, to probe how model predictions change under controlled perturbations (Fig. \ref{fig:hippo-schematic}b). HIPPO leverages the permutation-invariant nature of MIL models, which treat WSIs as unordered bags of patches, allowing these counterfactual patch sets to serve as valid inputs without requiring synthetic image generation or inpainting. Since MIL models are permutation-invariant, these resampled patch collections remain valid inputs to models. By enabling targeted interventions on patch sets, HIPPO allows us to systematically analyze how different histological features influence MIL model outputs. Unlike attention mechanisms, which highlight correlated regions but cannot test causal relationships, HIPPO supports hypothesis-driven analyses by explicitly measuring whether a region is necessary, sufficient, or irrelevant for a prediction. This capability elevates interpretability from descriptive observation to empirical hypothesis testing. Compared to traditional attention-based interpretations, HIPPO provides direct, quantitative assessments of the impact of specific tissue regions on model predictions.
}

We demonstrate how HIPPO enables our systematic understanding of the workings of a variety of pathology models performing clinical tasks: metastasis detection, prognostication, and \textit{IDH} mutation classification. First, in the context of metastasis detection, we evaluate six foundation models in pathology using the CAMELYON16 dataset \autocite{bejnordi2017camelyon16}. This dataset includes expert-annotated \rebuttal{tumor segmentations made by pathologists, providing ground-truth labels that allow us to rigorously determine whether regions deemed influential by the model correspond to established histopathologic ground truth.} \rebuttal{Throughout our metastasis detection experiments, we extensively use these expert-annotated tumor regions to validate HIPPO's causal attributions, testing both the necessity and sufficiency of tumor regions for model predictions against this ground truth.} Our analysis uncovers model-specific limitations and biases that would have remained hidden if relying solely on attention-based interpretations. We reveal that some models rely heavily on extratumoral tissue for metastasis detection, while others are surprisingly insensitive to small tumor regions. These findings highlight the importance of rigorous model evaluation beyond standard performance metrics and underscore the potential of HIPPO in identifying when and why models might fail. 
Second, building on this validation, we then applied HIPPO to cancer prognostication in breast cancer and melanoma datasets from The Cancer Genome Atlas (TCGA). Our results demonstrate that HIPPO can identify tissue regions more strongly associated with prognosis compared to those highlighted by attention. Strikingly, we find that high-attention regions can sometimes have counterintuitive effects on prognostic predictions, further emphasizing the limitations of relying solely on attention for model interpretation. By quantitatively assessing the impact of tumor-infiltrating lymphocytes (TILs) on model predictions, we confirm that the models have captured the known prognostic significance of TILs in both breast cancer and melanoma. This ability to link model behavior to established biological knowledge is crucial for building trust in AI-driven prognostic tools. Beyond model interpretation, HIPPO opens new possibilities for virtual hypothesis generation in computational pathology. 
Finally, we applied HIPPO to \textit{IDH} mutation classification in glioma and found that HIPPO outperformed attention in identifying the histological regions driving false negatives. By providing more precise and interpretable insights into model predictions, HIPPO enabled a clearer understanding of the reasons behind misclassifications.

As computational pathology continues to advance, the need for robust, interpretable, and trustworthy AI models becomes increasingly critical. \rebuttal{HIPPO addresses this need by providing a systematic framework for uncovering the strengths, limitations, and potential biases of models. By offering a more complete characterization of model behavior, HIPPO enhances the interpretability of existing approaches and facilitates the development of more reliable and clinically meaningful pathology models. As demonstrated across multiple applications, including metastasis detection, prognostic modeling, and mutation classification, HIPPO can help accelerate the translation of computational pathology methods into clinical practice and, ultimately, support improved patient care.}

\section*{Results}

\subsection*{HIPPO: Histopathology Interventions of Patches for Predictive Outcomes}

\rebuttal{HIPPO is a specimen-level perturbation framework that explains permutation-invariant MIL models in computational pathology (Fig. \ref{fig:hippo-schematic}a). The primary goal of HIPPO is to explore counterfactual scenarios (e.g., ``how would the model’s output change if the tumor microenvironment within a WSI were altered?'').} 
\rebuttal{Without relying on synthetic image generation, HIPPO performs such interventions through the exclusion or inclusion of single or multiple patches, leveraging the permutation-invariant structure of MIL models (Fig. \ref{fig:hippo-schematic}b). The resulting changes in model predictions can then be interpreted as counterfactual outcomes. Through this process, HIPPO provides quantitative insights into how specific tissue subregions might impact pathological assessments of the AI model. These assessments can include but are not limited to patient prognosis, treatment response prediction, metastasis detection, inference of spatial transcriptomics, gene mutation detection, and microsatellite instability identification. Applying HIPPO to pathology models enables researchers, regulators, and clinicians to elucidate model behavior and to better understand the relationship between patch-level tissue characteristics and model outputs.}

The key insight behind HIPPO lies in how data flows through MIL models. In these models, a WSI is treated as a bag of permutation-invariant patches, where the number and order of patches are allowed to vary. \rebuttal{This formulation enables interventions through either (1) removing specific patches, effectively excising tissue from the input specimen, or (2) including specific patches, simulating the addition of new tissue into the specimen (Fig. \ref{fig:hippo-schematic}b and Supplementary Fig. \ref{fig:hippo-schematic-insert-delete}). Since MIL models are essentially set functions, these modified patch collections remain valid model inputs and do not introduce spatial artifacts that could confuse models and confound interpretations (Supplementary Fig. \ref{fig:supp_permutation_invariance})}. This intervention approach enables us to explore hypothetical scenarios such as the exclusion or inclusion of tumor patches from a patient's specimen. Understanding how the predictions of MIL models change due to interventions provides quantitative insights into their decision making process, revealing important features and potential biases learned.

\rebuttal{Within the HIPPO framework, users can flexibly select which patches to include or exclude, guided either by prior hypotheses or by data-driven strategies. Accordingly, we categorize HIPPO usage into the following modes (see \nameref{sec:methods} and \nameref{sec:supplementary_methods} for details):}
\begin{itemize}
  \item \rebuttal{\textit{HIPPO-knowledge}: choosing patches to add or remove based on prior knowledge or a well-defined hypothesis and quantifying its effect by removing it from specimens or adding it to specimens without that region and measuring the change in model outputs. (Fig. \ref{fig:hippo-schematic}c). In other words, this hypothesis-driven approach assesses whether the areas considered to be important by experts truly affect model outputs.}
  \item \rebuttal{\textit{HIPPO-attention}: choosing a subregion to ablate based on the attention scores assigned by MIL models and quantifying its effect by removing it from specimens and measuring the change in model outputs (Fig. \ref{fig:hippo-schematic}d). In other words, this method assesses whether the areas with attention scores are indeed influential for the model's predictions.}
  \item \rebuttal{\textit{HIPPO-search-positive-effect}: a greedy search algorithm to identify the regions that most strongly drive a prediction to the positive direction. We apply this method to samples predicted as positive by the model to identify regions influential for their predictions (Fig. \ref{fig:hippo-schematic}e). This data-driven intervention can automatically identify important patches from scratch without relying on prior hypotheses.} 
  \item \rebuttal{\textit{HIPPO-search-negative-effect}: a greedy search algorithm to identify the regions that most strongly drive a prediction to the negative direction. We apply this method to samples predicted as negative by the model to identify regions influential for their predictions.}
\end{itemize}

With the advent of digital pathology foundation models, it is important to evaluate model robustness, generalizability, and potential biases and understand their limitations. Here, we showcase the advance  enabled by HIPPO in rigorously evaluating models built on top of foundation models for breast metastasis detection, prognosis prediction tasks, and IDH mutation detection task. We compare \rebuttal{six} foundation models in metastasis detection and identify model-specific limitations and biases. We also use HIPPO to study the effects of tissue components on prognostic models, demonstrating how HIPPO's capabilities surpass attention in identifying low and high-risk drivers. We also measure the effect of TILs on breast cancer and melanoma patient prognosis and demonstrate digitally that autologous TILs improve predicted prognosis in a subset of melanoma patients, marking exciting progress in the field.

\subsection*{Do MIL models think tumor is necessary for breast cancer metastasis detection?}

Metastasis detection is a well-studied task with well-defined features (i.e., tumor cells) that drive the label of whether or not a specimen contains metastasis. In a clinical setting, it is critical that metastases are identified --- a false negative is unacceptable. Recent studies have shown that MIL models have strong performance in metastasis detection\autocite{chen2024uni}. However, previous studies have also found that computer vision models can make the correct predictions for the wrong reasons, such as short-cut features or spurious correlations \autocite{geirhos2020shortcut, degrave2021shortcuts}. Thus, the degree to which AI models rely on the tumor regions remains to be seen, even for a relatively straightforward task like tumor detection. Understanding this is critical to elucidate the strengths and limitations of MIL models for metastasis detection, including potential biases.

\begin{figure}[ht!]
\centering
\includegraphics[width=7in]{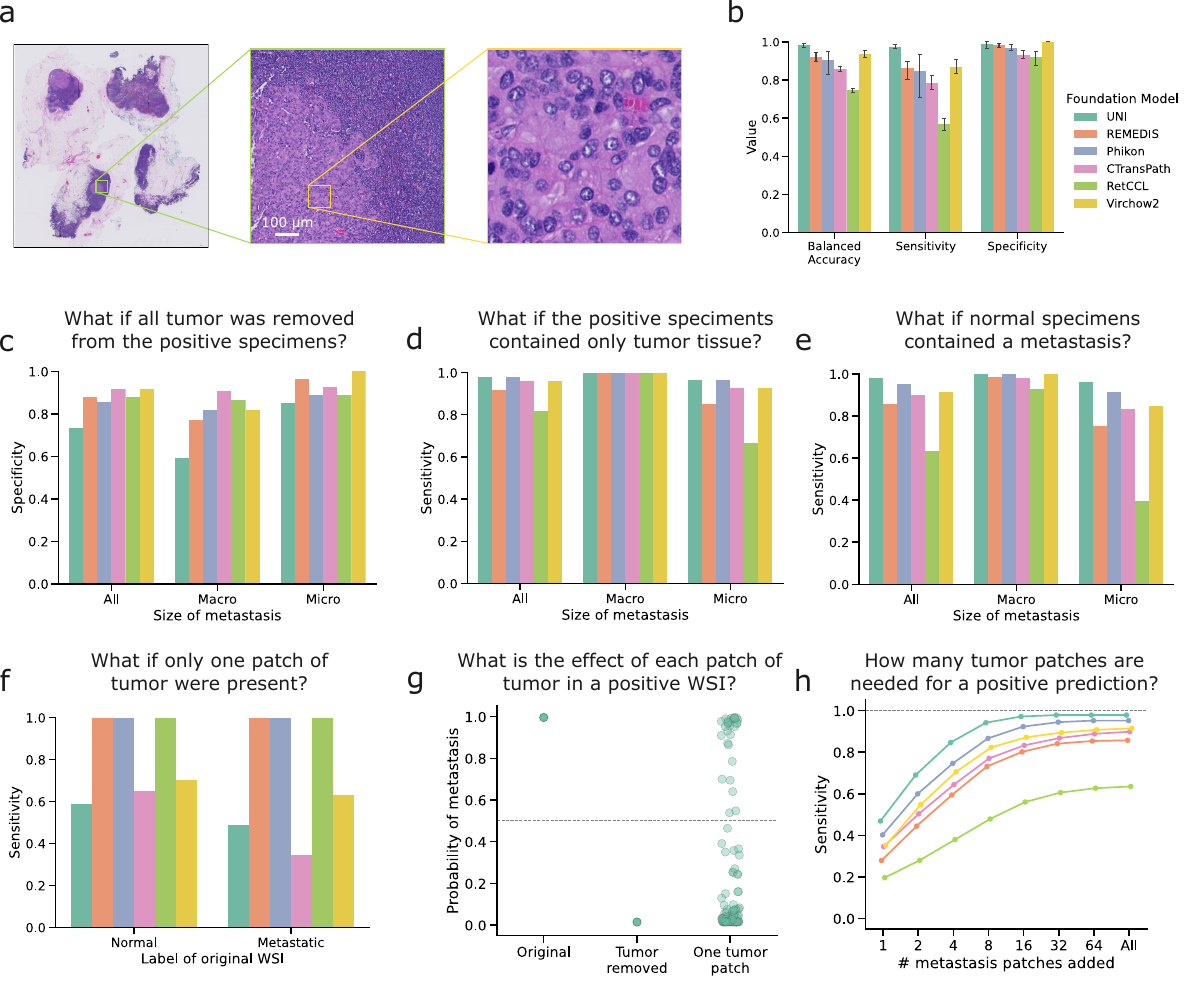}
\rebuttaltemp{Modified figure: Added results for Virchow2}
\caption{\small{\textbf{Understanding the role of tumor in detecting metastases.} \textbf{a,} Example WSI from the CAMELYON16 dataset containing a macrometastasis (specimen \texttt{test\textunderscore001}), with a $128\times128$ \SI{}{\micro\meter} patch highlighted. \textbf{b,} Bar plot of balanced accuracy, sensitivity, and specificity on the CAMELYON16 test set (n=129, 80 negative, 49 positive) across five random initializations and five encoders, with mean values and 95\% confidence intervals. The best-performing model for each encoder was used in subsequent experiments. \textbf{c-d,} Bar plots showing specificity when tumor-containing patches are removed (c) and sensitivity when only tumor tissue remains (d) in positive specimens (n=49, 22 macrometastases, 29 micrometastases), quantifying necessity and sufficiency of tumor regions for metastasis detection. \textbf{e,} Bar plot of sensitivity after adding metastases to negative specimens (3920 counterfactuals: 80 negative × 49 positive), further quantifying tumor sufficiency. \textbf{f}, Bar plot showing sensitivity of counterfactuals with a single $128 \times 128$ \SI{}{\micro\meter} tumor patch in normal (n=80) and metastatic (n=49) specimens. \textbf{g,} Strip plot of model probabilities for tumor patches in specimen \texttt{test\textunderscore051} using the UNI-based ABMIL model, comparing original, tumor-removed, and single-tumor-patch (n=125) conditions. \textbf{h,} Line plot relating tumor size to model sensitivity, with each point representing 3920 counterfactuals (80 negative × 49 positive) as tumor patches are added to negative specimens.
}}
\label{fig:cam16-nec-suff}
\end{figure}

To evaluate this, we trained several ABMIL models for breast metastasis detection using the CAMELYON16 dataset \autocite{bejnordi2017camelyon16} (Fig. \ref{fig:cam16-nec-suff}a and \nameref{sec:methods}). \rebuttal{While HIPPO is applicable to any permutation-invariant MIL models, we focus on ABMIL models because they are the de facto standard in computational pathology among MIL approaches.} Several pathology foundation models have recently emerged, demonstrating near-human levels in metastasis detection. Here we consider six pathology foundation models (UNI \autocite{chen2024uni}, REMEDIS \autocite{azizi2022remedis}, Phikon \autocite{filiot2023phikon}, CTransPath \autocite{wang2022ctranspath}, RetCCL \autocite{wang2023retccl}, and \rebuttal{Virchow2\autocite{vorontsov2024virchow, zimmermann2024virchow2})}. We trained five ABMIL models for each foundation model to distinguish whether or not a specimen contained metastasis. Similar to previously reported results \autocite{chen2024uni}, UNI achieved a mean balanced accuracy of 0.982, REMEDIS 0.922, Phikon 0.907, CTransPath 0.858, RetCCL 0.745, and \rebuttal{Virchow2 0.934} (Fig. \ref{fig:cam16-nec-suff}b, Supplementary Table \ref{table:supp_cam16_performance}).
For HIPPO explainability experiments, we used the best-performing model (out of 5 random initializations) on the test set for each foundation model. The best UNI model achieved balanced accuracy of 1.00, REMEDIS 0.949, Phikon 0.955, CTransPath 0.885, RetCCL 0.769, \rebuttal{and Virchow2 0.959} (Supplementary Table \ref{table:supp_cam16_performance_single_models}).

In this dataset, expert pathologists have produced high-resolution annotations of metastatic tumor regions. \rebuttal{These expert-annotated ground-truth labels allow us to use the \textit{HIPPO-knowledge} method to determine whether metastatic regions are necessary for detecting breast cancer metastasis, with the expert annotations serving as the ground-truth reference.} Specifically, for patients who were positive for metastasis, we created counterfactual examples by removing all patches that intersected with the tumor annotations. If we assume that a model must rely on tumor-containing patches to make a prediction, an accurate model would always return a negative prediction when all tumor containing patches have been removed. In the following, we use specificity to quantify model behavior. Specificity is defined as the ratio of true negatives to all negative samples. In this set of counterfactuals, all specimens were negative, so the specificity represented the proportion of correct negative predictions by the models. 

We compared model predictions before and after removing patches containing tumor (Fig. \ref{fig:cam16-nec-suff}c). Notably, the UNI-based model exhibited the lowest specificity (0.73) in these counterfactual examples despite achieving the highest balanced accuracy on the original test set (1.00). This discrepancy was particularly pronounced in counterfactual specimens that originally contained macrometastases (specificity 0.59), suggesting that the UNI-based ABMIL model uses tissue outside of the tumor region to drive positive metastasis predictions. The REMEDIS-based model exhibited a similar trend, with a specificity of 0.77 in counterfactuals derived from macrometastases. In contrast, the other models showed less dependence on extratumoral tissue (sensitivity of Phikon-based, 0.86; CTransPath-based, 0.92; RetCCL-based, 0.88; \rebuttal{Virchow2-based, 0.92}), indicating that their predictions are primarily driven by tumor epithelial cells rather than other tissue components.

We hypothesized that the low specificity of the UNI-based ABMIL model may be attributed to metastasis-induced alterations in the surrounding tumor microenvironment. To investigate this, we used \textit{HIPPO-knowledge} to remove increasingly larger regions surrounding the tumor annotation and quantified the effect on metastasis detection. As the extent of peritumoral tissue removal increased, the UNI-based model was consistently more likely to predict the absence of metastasis  (Supplementary Fig. \ref{fig:supp_nec_dilations}). Specificity increased from 0.73 at dilation of \SI{0}{\micro\meter}, to 0.78 at \SI{64}{\micro\meter}, to 0.80 at \SI{128}{\micro\meter}, to 0.86 at \SI{256}{\micro\meter}, and to 0.88 at \SI{1024}{\micro\meter}. This was driven primarily by macrometastatic specimens, where specificity increased from 0.59 at dilation of \SI{0}{\micro\meter} to 0.68 at dilation of \SI{64}{\micro\meter}, to 0.73 at dilation of \SI{128}{\micro\meter}, to 0.82 at dilation of \SI{256}{\micro\meter}, to 0.86 at dilation of \SI{1024}{\micro\meter}. Notably, other ABMIL models remained largely unaffected by peritumoral tissue removal, highlighting a unique characteristic of the UNI-based model. \rebuttal{These observations may reflect a dependence of the UNI-based model on microenvironmental alterations associated with metastasis, though they may also indicate that the model has learned a less robust boundary for identifying metastatic disease.}
In summary, HIPPO enabled the quantitative exploration of peritumoral tissue on metastasis detection.

\subsection*{Is tumor sufficient for breast cancer metastasis detection?}

While necessity assesses the importance of a feature or feature set, it does not inform whether the feature set is sufficient for model predictions. Metastasis detection models must be able to detect tumor regions no matter how small. Using the \textit{HIPPO-knowledge} method \rebuttal{with expert-annotated tumor regions as ground truth,} we tested the sufficiency of metastatic regions using two methods: removing all non-tumor patches and measuring model outputs and adding tumor regions to normal specimens and measuring model outputs. 

First, we constructed counterfactual specimens (n=49) by removing all non-tumor tissue (i.e., removing patches that did not intersect with expert tumor annotations) and measuring model outputs (Supplementary Fig. \ref{fig:hippo-schematic-insert-delete}a). Because these counterfactuals contained only tumor tissue, all images should be labeled ``positive.'' Under these conditions, the foundation models had the following sensitivity (true positive rate): UNI 0.98, REMEDIS 0.92, Phikon 0.98, CTransPath 0.96, RetCCL 0.82, \rebuttal{Virchow2 0.96} (Fig. \ref{fig:cam16-nec-suff}d). 

\rebuttal{By comparing these tumor-only predictions with those from the original WSIs, we assessed whether extratumoral tissue contributed to false negative errors. For micrometastatic specimens, five of the six foundation models showed improved sensitivity on the tumor-only images, suggesting that surrounding tissue influenced misclassification. Specifically, sensitivity increased by 25\% for CTransPath, 4\% for Phikon, 5\% for REMEDIS, 100\% for RetCCL, and 6\% for Virchow2}. For UNI, the original WSIs yielded a perfect sensitivity for micrometastasis. However, when evaluated on tumor-only images, UNI generated a single false negative, suggesting that the model relied on extratumoral tissue to make a positive prediction. In this specimen, tumor tissue alone was insufficient for UNI to classify the slide as metastatic, indicating that surrounding regions were solely driving the positive prediction.
A similar effect was observed with RetCCL in macrometastatic specimens: its true positive rate on the original WSIs was 0.95 (21 of 22 cases), whereas all macrometastases were correctly identified when evaluated on tumor-only images, demonstrating that extratumoral tissue caused the lone false negative in the original predictions.

We also evaluated whether tumor itself was sufficient for metastasis detection by embedding tumor regions into normal specimens (Supplementary Fig. \ref{fig:hippo-schematic-insert-delete}b). Specifically, we embedded all patches intersecting with tumor annotations into normal specimens, resulting in 3,920 positive counterfactual examples (80 normal slides $\times$ 49 positive slides). Model outputs were then recorded for these tumor-in-normal counterfactuals  (Fig. \ref{fig:cam16-nec-suff}e). The resulting sensitivities were: UNI-based 0.98, REMEDIS-based 0.86, Phikon-based 0.95, CTransPath-based 0.90, RetCCL-based 0.63, \rebuttal{and Virchow2-based 0.92}. Positive counterfactuals constructed with micrometastases were detected less reliably across most models (UNI-based achieved sensitivity of 0.96, REMEDIS-based 0.75, Phikon-based 0.91, CTransPath-based 0.93,  RetCCL-based 0.40, \rebuttal{Virchow2-based 0.85}), suggesting that smaller tumors in the context of normal tissue are insufficient for positive metastasis detection.

Finally, we quantified how strongly each tumor region generalizes across individuals by computing an average treatment effect for each metastatic slide.  For each positive slide, we averaged the model’s predicted probability of metastasis across all tumor-in-normal counterfactuals, providing a measure of how reliably that tumor induces a positive prediction across diverse normal backgrounds. Using this metric, 100\% of macrometastases (n=22) produced a positive average effect in the UNI-based, REMEDIS-based, Phikon-based, CTransPath-based, \rebuttal{and Virchow2-based} models. In the RetCCL-based model, 90\% (20 of 22) of macrometastases produced a positive average effect. Micrometastases (n=27) were less likely to induce positive predictions on average, with 96\% (n=26) positive in UNI, 93\% (n=25) in Phikon, 81\% (n=22) in CTransPath, 74\% (n=20) in REMEDIS, 37\% (n=10) in RetCCL, \rebuttal{and 85\% (n=23) in Virchow2}.

\subsection*{Foundation models may miss small breast cancer metastases}

\rebuttal{To evaluate how metastasis size affects the sensitivity of ABMIL models, we analyzed metastasis-positive specimens from the CAMELYON16 test set. Our approach first involved initially removing all patch embeddings intersecting with expert tumor annotations, thereby creating metastasis-removed versions of each positive slide. effectively rendering the slide negative for metastases. We then introduced a single $128 \times 128$ \SI{}{\micro\meter} tumor region (shown in the right-hand side of Fig. \ref{fig:cam16-nec-suff}a) into both 80 normal specimens and 49 metastasis-removed positive specimens. The REMEDIS-, Phikon-, and RetCCL-based ABMIL models detected 100\% of counterfactuals as positive, demonstrating robustness to detecting even very small signature of tumor (Fig. \ref{fig:cam16-nec-suff}f).  In contrast, the UNI-based model failed to detect 41\% (33 of 80) of positive counterfactuals, the CTransPath-based models failed 35\% (28 of 80), and the Virchow2-based models failed 30\% (24 of 80). A similar pattern was observed when the tumor region was embedded into the metastasis-removed positive specimens: REMEDIS-, Phikon-, and RetCCL-based models detected 100\% of counterfactuals (49 of 49), while the UNI-based model missed 51\% (25 of 49), the CTransPath-based model missed 65\% (32 of 49), and the Virchow2-based model missed 37\% (18 of 49) (Fig. \ref{fig:cam16-nec-suff}f). This result is unexpected because the UNI-based model exhibited perfect sensitivity on the original CAMELYON16 test set (Fig. \ref{fig:cam16-nec-suff}b) and achieved the highest sensitivity when larger tumors were embedded into normal tissue (Fig. \ref{fig:cam16-nec-suff}e). This highlights that strong classification performance on the held-out test set does not necessarily reflect a model’s ability to generalize to more fine-grained or clinically relevant downstream applications.}

We also sought to quantify how sensitive each model is to individual tumor patches within positive specimens, as this can reveal whether different regions of tumor carry different levels of informativeness for machine learning classifiers. To accomplish this, \rebuttal{we first removed all tumor patches intersecting with expert tumor annotations, producing a tumor-removed specimen. We then reintroduced patches that were fully contained within the expert tumor annotation, one at a time, into the tumor-removed specimen and evaluated the model outputs. These outputs were compared with the predictions made on the fully tumor-removed slide.  While some individual tumor patches were sufficient to  drive a positive prediction on their own, many were not (Fig. \ref{fig:cam16-nec-suff}g for the UNI-based model, and other models and specimens are shown in Supplementary Figs. \ref{fig:supp_all_single_patches_uni}-\ref{fig:supp_all_single_patches_virchow2}).}

\rebuttal{To further quantify how tumor size influences metastasis detection, we progressively added increasing numbers of tumor patches into normal slides and measured model sensitivity. All models demonstrated a graded response to tumor size,  with sensitivity increasing as more tumor was added  (Fig. \ref{fig:cam16-nec-suff}h). UNI exhibited the highest sensitivity. For most models, sensitivity plateaued after approximately \SI{0.262}{\milli\meter\squared} of tumor (16 patches) had been added. The RetCCL-based model showed the lowest overall sensitivity and was the least responsive to smaller tumor regions.}

\rebuttal{To identify the largest amount of tumor that could be added while still going undetected, we applied a variant of \textit{HIPPO-search-negative-effect}. Whereas the original \textit{HIPPO-search-negative-effect} removes patches within a WSI, this variant progressively adds tumor patches to the ``negative counterfactuals'' described above. For each iteration, we selected the tumor patch whose effect size is lowest using the $\arg\min$ function. Using this procedure, we found that in some cases, regions up to \SI{1.5}{\milli\meter\squared} could be added into a negative counterfactual without triggering a positive metastasis prediction. The tumor patches with the low effect sizes were largely similar in appearance to those that were sufficient to induce positive predictions, although some low-effect patches contained adipose cells along with tumor epithelial cells (Supplementary Fig. \ref{fig:supp_mets_individual_tumor_patches_unseen}). 
This shows that substantial regions of tumors can remain undetected by an ABMIL-based metastasis detection model. These biases warrant further investigation before these models are deployed in clinical settings.}

\subsection*{Peritumoral tissue can cause false positive metastasis detections}

Given the influence of peritumoral tissue on UNI-based model predictions, we next evaluated whether peritumoral tissue were themselves sufficient to drive positive metastasis predictions. From each metastasis-positive specimen (n=49), we extracted halos of peritumoral tissue with widths of 64, 128, 256, or 1024 \SI{}{\micro\meter}. Halos were defined beginning at either at the boundary of the expert tumor annotation or \SI{256}{\micro\meter} outside of the tumor annotation. Patches intersecting these halos were then added to normal specimens (n=80), resulting in 3,920 counterfactual examples (80 normal $\times$ 49 positive specimens). Model predictions were averaged across the normal specimens to compute the average treatment effect for each peritumoral region. This was evaluated for the UNI-based and Phikon-based models.

Halos of peritumoral tissue alone were sufficient to drive positive metastasis detection. Halos \SI{1024}{\micro\meter} in width beginning at the tumor boundary caused positive predictions in 20\% (n=10) of positive specimens and 10\% (n=5) in the Phikon-based model (Supplementary Fig. \ref{fig:supp_peritumoral}). When the \SI{1024}{\micro\meter} halo \SI{256}{\micro\meter} was shifted \SI{256}{\micro\meter} outside the tumor boundary, positive predictions occurred in 14\% (n=7) of specimens for the UNI-based model and 10\% (n=5) for the Phikon-based model. Even thin halos of \SI{64}{\micro\meter} beginning \SI{256}{\micro\meter} outside the tumor boundary triggered positive calls in 10\% (n=5) of UNI-based cases and 6\% (n=3) of Phikon-based cases (Supplementary Fig. \ref{fig:supp_peritumoral}). These results demonstrate that both models learned association between peritumoral tissue and metastasis status, despite the absence of any metastatic cells in these regions. Importantly, this bias could not have been uncovered using attention alone. HIPPO enabled the quantitative assessment of the influence of peritumoral tissue on metastasis detection across multiple foundation models.

\subsection*{\rebuttal{HIPPO can help confirm regions that lead to false negative metastasis detections}}

\begin{figure}[htbp]
\centering
\includegraphics[width=6.5in]{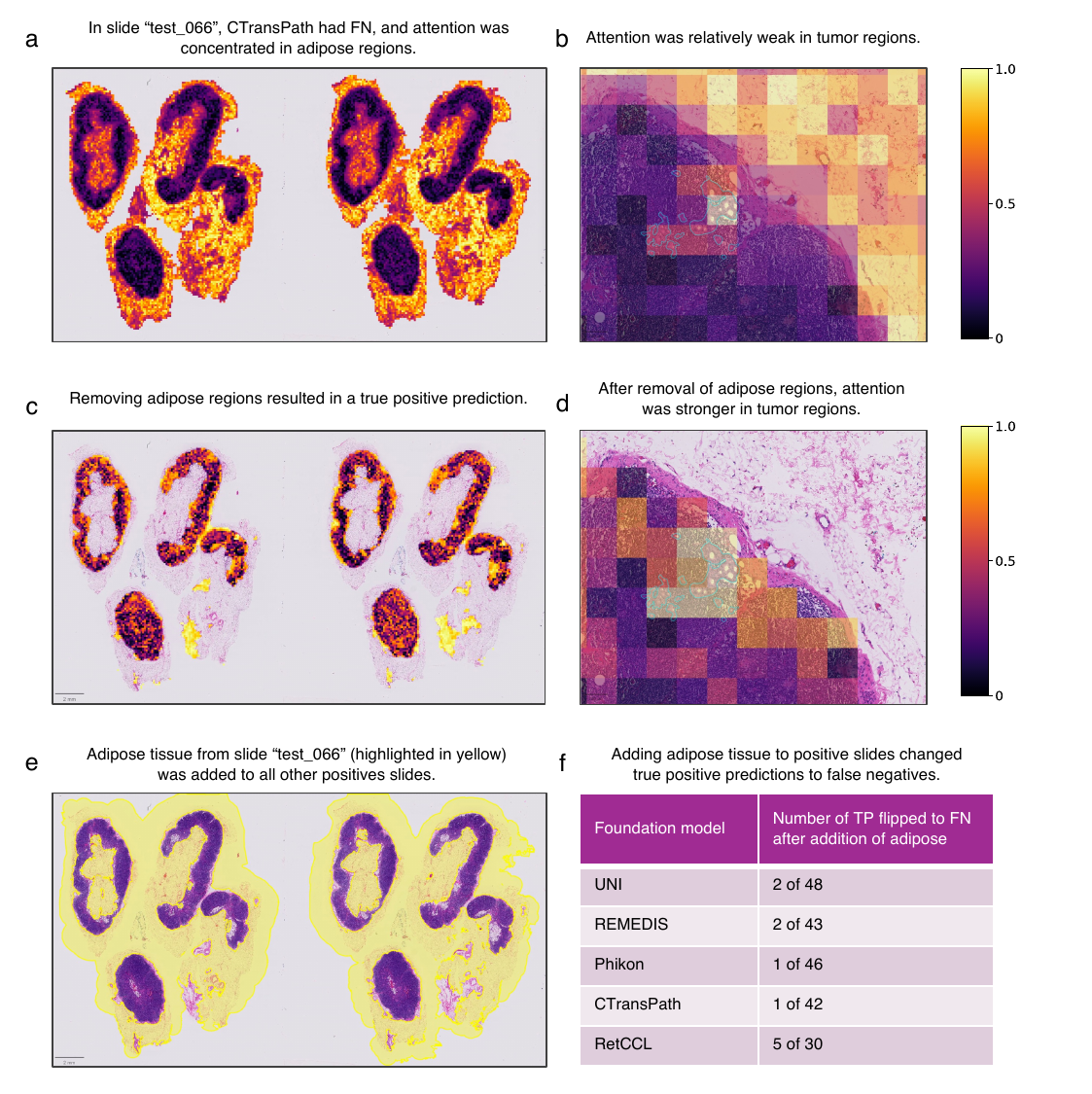}
\caption{\small{\rebuttal{\textbf{HIPPO can help confirm regions that lead to false negative metastasis detections.} As a demonstration of HIPPO's capability to diagnose model errors}, we investigated a case where the CTransPath-based ABMIL model misclassified specimen \texttt{test\textunderscore066} as negative. \textbf{a-b,} Attention was strongest in adipose regions and relatively weak in the tumor. This led us to hypothesize that the adipose regions were causing the misclassification. \textbf{c-d,} The true positive prediction was rescued after adipose regions were removed. In addition, attention was concentrated in tumor regions (\textbf{d}). To evaluate how general the effect of this adipose tissue was on preventing true positive predictions, we added the embeddings from the adipose region (yellow highlighted region in \textbf{e}) to all positive slides except \texttt{test\textunderscore066}. \textbf{f,} The addition of adipose tissue caused several true positive predictions to switch to false negatives, suggesting that the adipose tissue may cause false negatives. \rebuttal{This demonstrates how HIPPO can help confirm which regions of a WSI are causing misclassifications, enabling targeted diagnosis of model errors.}}}
\label{fig:cam16-adipose}
\end{figure}

To further investigate patient samples that consistently produced false negative predictions, we visualized the attention maps of the ABMIL models to identify regions considered important by the model. In the CTransPath-based model, attention for specimen \texttt{test\textunderscore066} was primarily focused on adipose regions  (Fig. \ref{fig:cam16-adipose}a-b). Since attention maps only provide a qualitative visualization of regions in an image that the ABMIL models consider important, it is unclear to what extent adipose tissue directly affects model predictions. We address this with \textit{HIPPO-attention} (\nameref{sec:methods}). \rebuttal{The adipose regions were annotated and verified by an expert pathologist to ensure accurate identification.} The adipose patches were then removed, and the effect of this perturbation on model outputs was quantified. Removing the adipose tissue restored the true positive prediction in specimen \texttt{test\textunderscore066}, \rebuttal{demonstrating that HIPPO can confirm which regions of a WSI are responsible for misclassification} (Fig. \ref{fig:cam16-adipose}c-\ref{fig:cam16-adipose}d). 

To test whether the adipose tissue from \texttt{test\textunderscore066} could also induce misclassification in other models and specimens, we added the adipose regions from that specimen into each of the 48 remaining metastasis-positive specimens and recorded whether the added adipose caused false negatives (Supplementary Fig. \ref{fig:hippo-schematic-insert-delete}b). We found that true positives were flipped to false negatives in 2, 2, 1, 1, and 5 specimens for UNI, REMIDIS, Phikon, CTransPath, and RetCCL, respectively (Fig. \ref{fig:cam16-adipose}e-\ref{fig:cam16-adipose}f). This case study illustrates how HIPPO can elucidate biases that drive misclassification. While attention alone could not determine whether adipose tissue caused misclassification, it provided a useful hypothesis that we could formally evaluate with HIPPO-attention. This illustrates how HIPPO can complement attention-based interpretability analysis by quantitatively testing whether putatively important tissue regions truly influence model predictions.

\subsection*{HIPPO identifies shortcut learning when attention struggles}

Identifying spurious correlations in deep learning models for medical imaging is essential for ensuring reliable and clinically meaningful predictions. To test HIPPO's ability to identify spurious correlations,  we deliberately introduced an artificial bias into the CAMELYON16 dataset (Supplementary Figs. \ref{fig:cam16-shortcuts}a and \ref{fig:cam16-shortcuts}b). Specifically, we inserted $768 \times 768$ \SI{}{\micro\meter} blue squares into all negative images, mimicking a plausible real-world scenario in which a pathologist marks certain slides with a blue marker. This modification creates a strong spurious correlation between the presence of a blue region and the negative  label. We hypothesized that a model trained on this modified dataset would rely on the presence or absence of the blue region, rather than tumor morphology, to infer slide level labels. 

An ABMIL model using UNI embeddings was trained on the modified dataset and achieved a balanced accuracy of 1.0 on the test set, indicating that the spurious correlations turned the task into a trivial classification problem. By performing standard model interpretation using attention,  we found that metastatic regions were considered highly important (Supplementary Fig. \ref{fig:cam16-shortcuts}c). However, removing these regions using HIPPO did not alter the model predictions, demonstrating that tumor regions were not important for model predictions despite a strong attention assignment. This highlights an important weakness of attention:  the regions with high attention weights may not be the regions that actually drive predictions.

Knowing that metastatic regions did not affect the model's outputs, we used the search algorithm \textit{HIPPO-search-positive-effect} (\nameref{sec:methods}) to identify the patches that maximally drove positive tumor predictions in both models for one positive specimen, \texttt{test\textunderscore002}. Given that the model trained with spurious correlations uses the lack of a blue square as a cue for positive specimens, we expected that no individual patches would strongly drive the positive metastasis output and that tumor regions would not have a high effect on the prediction. Indeed, effect sizes were small and evenly distributed across the WSI (minimum \SI{2.1e-05}{}, maximum \SI{0.02}{}, mean \SI{9.4e-05}{}, and median \SI{5.5e-05}{}), indicating that no single region contributed strongly to the model prediction (Supplementary Fig. \ref{fig:cam16-shortcuts}b). By contrast, applying the same search algorithm to the model trained on the original CAMELYON16 datasetrevealed higher effect sizes (minimum \SI{3.7e-08}{}, maximum \SI{0.09}{}, mean \SI{1.3e-4}{}, and median \SI{4.9e-08}{}), with high-effect patches localized within expert tumor annotations (Supplementary Fig. \ref{fig:cam16-shortcuts}e). \rebuttal{We repeated this analysis across five additional positive specimens and consistently observed the same trends (Supplementary Table \ref{table:supp_spurious}). } 

\rebuttal{These results show that simply inspecting performance and attention maps can obscure major failure modes. Although the model with the artificial shortcut achieved perfect accuracy and highlighted tumor regions in its attention map, these signals were misleading: tumor patches had almost no causal influence on the prediction. In contrast, HIPPO quantified the effect sizes of individual regions and correctly revealed that the model’s decisions were driven entirely by the introduced bias rather than by tumor morphology. By grounding interpretation in measured changes to model predictions, HIPPO provides a more faithful view of model behavior and can uncover hidden shortcut learning that conventional observational methods fail to detect.}

\subsection*{Refining the search for prognostic tissue biomarkers}

\begin{figure}[!b]
\centering
\includegraphics[width=7in]{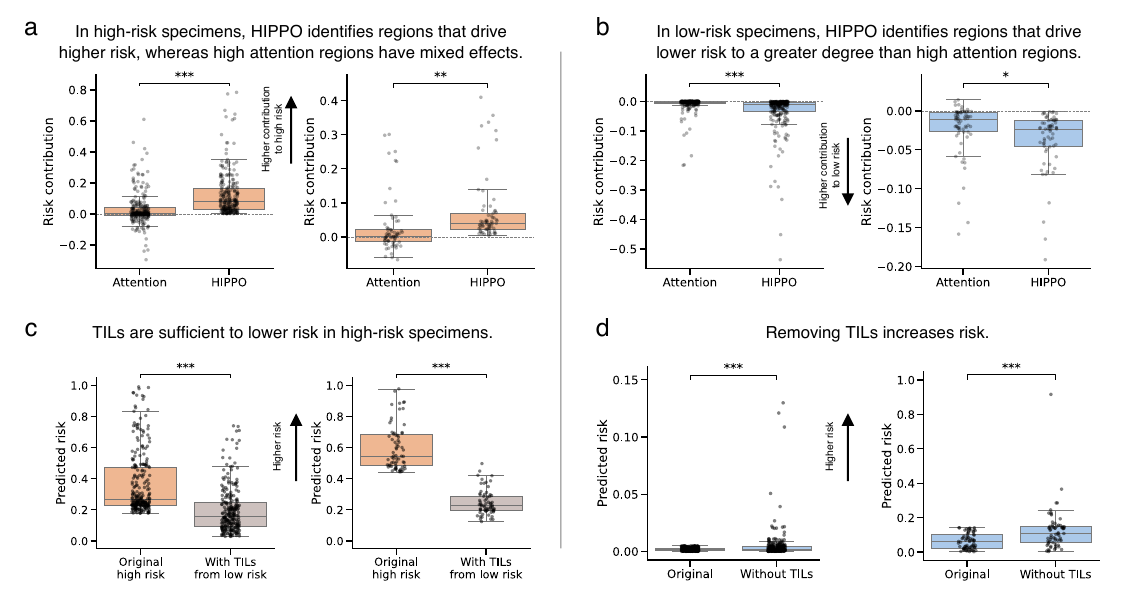}
\caption{\small{\textbf{HIPPO outperforms attention in identifying prognostic tissue regions.} We studied prognostic ABMIL models in invasive breast carcinoma (BRCA) and cutaneous melanoma (SKCM) from The Cancer Genome Atlas. 
\textbf{a, b,} Box plots of the prognostic effects of patches selected using attention and HIPPO in high-risk (\textbf{a}) and low-risk (\textbf{b}) specimens. The $y$-axis depicts the risk contribution, which is calculated as the original predicted risk minus the predicted risk when using a specimen with high-attention or high-HIPPO patches removed. Positive values indicate contribution to higher risk (\textbf{a}), and negative values indicate contribution to lower risk (\textbf{b}). The $x$-axis is the method of patch selection (either the top 1\% of attended patches or the top 1\% of patches found using \textit{HIPPO-search-positive-effect} or \textit{HIPPO-search-negative-effect}). 
\textbf{c,} Box plots showing the predicted risk scores before and after adding tumor-infiltrating lymphocytes (TILs) to high-risk BRCA (left, n=256) and SKCM (right, n=67) specimens. Orange boxes show the original risk scores, and gray boxes show risk scores after adding TILs from low-risk specimens and averaging across low-risk specimens. Lower risk scores indicate improved prognosis.
\textbf{d,} Box plots showing the predicted risk scores before and after removing TILs from low-risk BRCA (left, n=256) and SKCM (right, n=67) specimens.
Box plots show the first and third quartiles, the median (central line) and the range of data with outliers removed (whiskers), and significance is shown (*: $p < 0.05$, **: $p < 0.01$, ***: $p < 0.001$).
Sample sizes in high-risk (\textbf{a, c}) and low-risk (\textbf{b, d}) are n=256 for BRCA (left) and n=67 for SKCM (right).
}}
\label{fig:surv-results}
\end{figure}

Having demonstrated HIPPO's effectiveness in metastasis detection, where the regions of interest are well-defined and were previously annotated by expert pathologists, we extended our investigation to the more challenging task of cancer prognosis. Unlike metastasis detection, prognostic factors in WSIs are multifaceted and less clearly defined. We applied HIPPO to prognostic models that generate risk scores from WSIs, with the goal of identifying the tissue regions most responsible for these predictions. 

Our experiments yielded two key insights. First, HIPPO's search algorithms demonstrated superior ability in identifying tissue patches that consistently and significantly influence risk predictions compared to conventional attention. Attention-based interpretations often produced mixed effects or counterintuitive effects, such as highlighting regions that lowered predicted risk in otherwise clearly high-risk specimens. HIPPO, on the other hand, provided a consistent, reliable, and quantitative assessment of the regions that drive risk. Second, HIPPO's perturbation-based design enabled ``virtual experiments'' to assess the effect of targeted tissue interventions on prognostic outcomes in silico through the lens of the ABMIL model. This capability positions HIPPO as a promising tool for accelerating the discovery and validation of prognostic tissue biomarkers, thereby narrowing the gap between computational analysis and clinical actionability.

We trained prognostic ABMIL models using the PORPOISE framework \autocite{chen2022porpoise}, a computational tool designed for predicting survival outcomes from histopathology images, to predict overall survival from WSIs in breast cancer (TCGA-BRCA) and cutaneous melanoma (TCGA-SKCM). The same training and validation splits were used as in the original publication. Non-overlapping $128 \times 128$ \SI{}{\micro\meter} patches from WSIs were embedded using the UNI model \autocite{chen2024uni} (in the original PORPOISE publication, a truncated ResNet50 \autocite{he2016resnet} was used). \rebuttal{The resulting models successfully stratified patients into low- and high-risk groups, achieving concordance indices of 0.667 and 0.557 for TCGA-BRCA and TCGA-SKCM, respectively (Supplementary Fig. \ref{fig:supp_surv_km}).}  Low and high risk were defined as the first and fourth quartiles of risk scores. High attention regions were defined as the top 1\% of attended patches. For comparison, HIPPO search algorithms was used to identify the top 1\% of patches based on effect size.

High attention regions drove counterintuitive effects in many specimens, while \textit{HIPPO-search-positive-effect} and \textit{HIPPO-search-negative-effect} identified more robust and consistent drivers of risk. High attention regions in high-risk cutaneous melanoma specimens (n=67) drove lower risk in 45\% (n=30) of specimens. \textit{HIPPO-search-positive-effect}, on the other hand, identified regions that all drove higher risk and that more greatly contributed to high-risk predictions ($t=3.03$, $p<0.01$, independent two-sided t-test). High attention in high-risk breast cancer specimens (n=256) drove lower risk in 40\% (n=102) specimens. Again, \textit{HIPPO-search-positive-effect} consistently identified regions that drove higher risk in the high-risk specimens ($t=8.83$, $p<0.0001$, independent two-sided t-test) (Fig. \ref{fig:surv-results}a). High attention regions in low-risk SKCM specimens (n=67) drove higher risk in 10\% (n=7). \textit{HIPPO-search-negative-effect} identified regions that all drove lower risk and more strongly contributed to lower risk predictions ($t=-2.30$, $p<0.05$, independent two-sided t-test). High attention regions in low-risk BRCA specimens (n=256) drove higher risk predictions in 8\% (n=20) specimens. \textit{HIPPO-search-negative-effect} identified patches that consistently drove lower risk predictions ($t=-5.43$, $p<0.0001$, independent two-sided t-test) (Fig. \ref{fig:surv-results}b). These counterintuitive effects underscore that attention scores do not necessarily correspond to a region’s influence on the model output, and that interpretations relying solely on attention may be misguided. In contrast, HIPPO search algorithms robustly identified the regions that drove risk predictions and may be valuable tools for discovering prognostic biomarker.

\rebuttal{TILs are a well-established prognostic biomarker. We evaluated their necessity and sufficiency for low-risk predictions in BRCA and SKCM using HIPPO. To test sufficiency, we extracted TIL-positive patches from low-risk specimens and placed them in high-risk specimens (Supplementary Fig. \ref{fig:hippo-schematic-insert-delete}b). For each high-risk slide, we averaged predictions across TILs sourced from low-risk slides to compute the average treatment effect. In high-risk BRCA specimens (n=253, three specimens failed cell detection), adding TILs from low-risk specimens decreased the risk by 46\% ($t=17.95, p < 0.0001$, paired two-sided t-test) from 0.37 (std. dev. 0.20) to 0.20 (std. dev. 0.15). In SKCM (n=67), TIL addition significantly decreased risk by 59\% ($t=-22.53, p<0.0001$, paired two-sided t-test) from 0.60 (std. dev. 0.14) to 0.25 (std. dev. 0.08) (Fig. \ref{fig:surv-results}c). }

\rebuttal{To evaluate the necessity of TILs, we removed TIL-positive patches from low-risk specimens and measured the change in predictions (Supplementary Fig. \ref{fig:hippo-schematic-insert-delete}a). If TILs are necessary for low-risk predictions, risk should increase upon removal. In BRCA (n=254, two excluded), removaling TILs increased risk by 179\% ($t=3.83, p<0.001$, paired two-sided t-test) from 0.002 (std. dev. 0.001) to 0.005 (std. dev. 0.014). In SKCM (n=67), risk increased by 98\% ($t=4.27, p<0.0001$, paired two-sided t-test) from 0.064 (std. dev. 0.045) to 0.126 (std. dev. 0.123) (Fig. \ref{fig:surv-results}d). Although removing TILs increased risk, the resulting risk predictions did not reach the level of high-risk slides, suggesting that other features in the WSIs were also driving the low-risk predictions. Together, these analyses show that HIPPO facilitates a quantitative evaluation of the role of prognostic features, such as TILs, providing insights not achievable through attention alone. }

\subsection*{Generating hypotheses of which patients may benefit from autologous TIL therapy}

Lifileucel is a promising immunotherapy for melanoma that involves isolating TILs from a patient's tumor, expanding them ex vivo, and reinfusing them  into the patient\footnote{\url{https://www.fda.gov/news-events/press-announcements/fda-approves-first-cellular-therapy-treat-patients-unresectable-or-metastatic-melanoma}}. In a phase II clinical trial, more than 30\% of patients responded to the treatment \autocite{sarnaik2021lifileucel}. Identifying which patients are most likely to benefit from this therapy could meaningfully improve patient outcomes and reduce costs, given that a single treatment may cost over \SI{500000}[\$]{} \autocite{healey2024lifileucelnews}). Motivated by this, we investigated whether HIPPO could be used to generate hypotheses about patient-specific responses using virtual experiments. 

We used the prognostic ABMIL model for cutaneous melanoma described above, and we studied the high-risk specimens from TCGA-SKCM (n=67 WSIs, n=54 patients). Counterfactuals were constructed to emulate autologous TIL infusion: in each specimen, TIL-positive patches were replicated $2 \times$, $10 \times$, $20 \times$, and $100 \times$ (Supplementary Fig. \ref{fig:supp_surv_autologous_tils}a). TIL-positive patches were defined using the heuristic described in \nameref{sec:methods}. Changes in model-predicted risk between original specimen and each counterfactual was recorded, and Cohen's \textit{d} was calculated to quantify effect sizes. Importantly, we emphasize that these experiments are not intended to evaluate the true efficacy of autologous TIL therapy, but rather to illustrate how HIPPO can be used for hypothesis generation.

Autologous TIL augmentation reduced predicted risk in a dose-dependent manner. Risk decreased by \SI{-2.18}{\%} ($d=-0.50$) at $2 \times$ dose ($t=-4.06$, $p<0.001$, paired two-sided t-test), \SI{-10.8}{\%} ($d=-0.56$) at $10 \times$ dose ($t=-4.59$, $p<0.0001$, paired two-sided t-test), \SI{-15.3}{\%} ($d=-0.62$) at $20 \times$ dose ($t=-5.06$, $p<0.0001$, paired two-sided t-test), and \SI{-20.8}{\%} ($d=-0.67$) at $100 \times$ dose ($t=-5.49$, $p<0.0001$, paired two-sided t-test) (Supplementary Fig. \ref{fig:supp_surv_autologous_tils}b). Increasing the number of TILs by $100 \times$ decreased predicted risk scores by over half in \SI{18}{\%} of high-risk specimens. Together, these results demonstrate a proof-of-principle that HIPPO can be used to generate hypotheses about which patients may derive prognostic benefit from autologous TIL therapy, by simulating the effect of TIL replication on patient-specific risk. 

\subsection*{HIPPO addresses limitations of attention in explaining mutation detection}

\begin{figure}[!t]
\centering
\includegraphics[width=\linewidth]{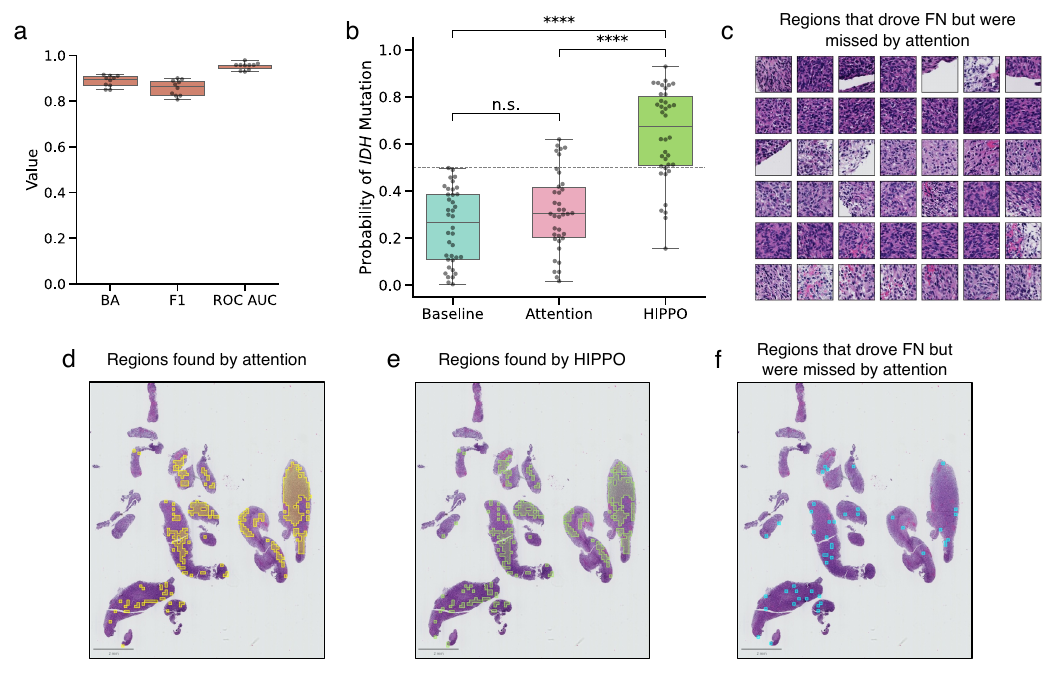}
\caption{\small{\textbf{HIPPO outperforms attention in identifying regions that cause misclassifications.}
\textbf{a,} Box plot showing model performance for IDH mutation classification, with 10-fold Monte Carlo cross validation. Balanced accuracy (BA), F1-score (F1), and the receiver operating characteristic area under the curve (ROC AUC) are depicted.
We sought to investigate the causes of false negative (FN) predictions (i.e., specimens which have an \textit{IDH} mutation but were classified as wildtype) (n=38 specimens). We compared the ability of attention and HIPPO to identify the regions that drove FNs. We removed the top 20\% of patches by attention and by \textit{HIPPO-search-negative-effect} (\nameref{sec:methods}). The HIPPO search found patches that increased probability of \textit{IDH} mutation when removed. 
\textbf{b,} Box plot of model probabilities in FN specimens at baseline (original specimens), attention, and HIPPO. Removing the patches found by HIPPO rescued true positive predictions in many cases and led to significantly higher model probabilities than baseline ($p<0.0001$, independent two-sided t-test) and attention ($p<0.0001$, independent two-sided t-test). Removing the top 10\% of patches by attention did not significantly change model predictions from baseline ($p>0.05$, independent two-sided t-test). Classification threshold is shown at $y=0.5$.
\textbf{c,} Sample of regions that HIPPO identified as driving FN predictions. These regions were identified by HIPPO but were not identified by attention.
FN specimen with \textbf{(d)} top 20\% of patches by attention, \textbf{(e)} top 20\% of patches by HIPPO, and \textbf{(f)} patches found by HIPPO but not found by attention. While the regions found by attention and HIPPO may appear similar, the differences are enough to mean the difference between false negative and true positive.
Box plots show the first and third quartiles, the median (central line) and the range of data with outliers removed (whiskers), and significance is shown (****: $p < 0.0001$, n.s.: not significant).
}}
\label{fig:ebrains-fn}
\end{figure}

To evaluate HIPPO’s advantages over attention in diagnosing misclassifications, we applied it to \textit{IDH} mutation classification, a prognostic marker in gliomas \autocite{han2020idh}. Using ABMIL models trained on the EBRAINS dataset \autocite{roetzer2022digital} (n=873), we classified wildtype (n=540) and mutant (n=333) \textit{IDH} in H\&E-stained specimens, employing 10-fold Monte Carlo cross-validation to assess variability. The classifiers demonstrated strong performance, achieving a balanced accuracy of $0.89 \pm 0.03$ (mean $\pm$ std. dev.) (Fig. \ref{fig:ebrains-fn}a), although false negative (FN) rates varied widely across folds (0.03-0.26). We used HIPPO and attention to identify and remove influential patches in FN specimens, with the goal of increasing the model’s \textit{IDH} mutation probability.

Removing the top \SI{20}{\%} of patches identified by \textit{HIPPO-search-negative-effect} significantly increased predicted \textit{IDH} mutation probabilities from baseline ($p<0.0001$, independent two-sided t-test, Cohen's $d=2.18$), converting \SI{76}{\%} of FNs into true positives. In contrast, removing the top \SI{20}{\%} of high-attention patches converted only \SI{16}{\%} of FNs, producing a nonsignificant change in mutation probability ($p>0.05$, independent two-sided t-test, Cohen's $d=0.38$) (Fig. \ref{fig:ebrains-fn}b). These results indicate that HIPPO is more effective than attention at identifying critical regions responsible for FN predictions.

While there was substantial overlap between the patches identified by HIPPO and attention (median: \SI{70}{\%}, range: \SI{28}{\%} to \SI{93}{\%}), the patches uniquely identified by HIPPO were often critical for correcting FNs (Fig. \ref{fig:ebrains-fn}c). For instance, in specimen \texttt{a1982c61-357f-11eb-b540-001a7dda7111}, occluding high-HIPPO patches increased the predicted \textit{IDH} mutation probability from 0.24 to 0.93 (Fig. \ref{fig:ebrains-fn}d), whereas occluding high-attention patches increased it only to 0.30 (Fig. \ref{fig:ebrains-fn}e). Despite an \SI{89}{\%} overlap between the two patch sets, the HIPPO-exclusive patches were essential for converting this FN to a true positive (Fig. \ref{fig:ebrains-fn}e).

HIPPO also demonstrated clear advantages over attention in identifying regions that drive positive predictions. For specimens predicted as \textit{IDH}-mutant, applying \textit{HIPPO-search-positive-effect} consistently revealed patches with stronger contributions to the positive prediction. Removing these HIPPO-identified patches produced more pronounced decreases in mutation probability than removing high-attention patches (Supplementary Figs. \ref{fig:supp_ebrains_pos}a and \ref{fig:supp_ebrains_pos}c).

Heatmaps derived from HIPPO outputs exhibited sharper delineation of critical regions compared to attention maps, potentially improving interpretability. Notably, HIPPO sometimes highlighted regions distinct from those with high attention scores, implying that diagnostically relevant areas may not always align with high-attention regions (Supplementary Figs. \ref{fig:supp_ebrains_pos}b and \ref{fig:supp_ebrains_pos}d). To quantify patch quality, we measured how many patches needed to be removed to reduce positive predictions below a probability threshold of 0.4. HIPPO required significantly fewer patches than attention ($p<0.001$, independent two-sided t-test), indicating that HIPPO identified patches with a stronger causal influence on model predictions. 

\subsection*{\rebuttal{HIPPO is generally applicable to any permutation-invariant MIL models beyond ABMIL}}

\rebuttal{While HIPPO is a versatile toolkit applicable to any permutation-invariant MIL models, our primary experiments concentrated on ABMIL architectures due to their status as the de facto standard in computational pathology. However, since alternative MIL architectures, particularly those based on transformer self-attention, are emerging, we also experimentally demonstrate that HIPPO generalizes beyond ABMIL.}  

\rebuttal{To this end, we extended the \textit{IDH} mutation analysis to a vision transformer (ViT)-based MIL model \autocite{dosovitskiy_image_2021, wagner2023transformer, el2025stamp}. The original ViT model \autocite{dosovitskiy_image_2021} was adapted to the MIL setting by using patch embeddings as tokens and removing positional embeddings to ensure permutation invariance \autocite{wagner2023transformer, el2025stamp}. Following the evaluation setup in Figure \ref{fig:ebrains-fn}b, we assess if HIPPO provides more accurate identification of influential patches than ViT's intrinsic attention mechanism (Supplementary Fig. \ref{fig:supp_idh_prediction_with_vit}). Specifically, we compared HIPPO-search-negative with two scores derived from ViT's attention mechanism, namely \textit{attention last} and \textit{attention rollout} \autocite{abnar2020quantifying} (see \nameref{sec:methods}).}

\rebuttal{Removing the top \SI{20}{\%} of patches identified by \textit{HIPPO-search-negative-effect} significantly increased predicted \textit{IDH} mutation probabilities from baseline ($p<0.0001$, independent two-sided t-test, Cohen's $d=3.45$), resulting in a true positive classification for \SI{82}{\%} of cases. In comparison, removing the top \SI{20}{\%} of ViT-based attention patches converted only \SI{18}{\%} of FNs to true positives when using \textit{attention last} and \SI{27}{\%} when using ViT attention rollout, with both producing nonsignificant effects on mutation probability ($p>0.05$, independent two-sided t-test, Cohen's $d=0.19$ and $0.60$, respectively). These findings show that HIPPO generalizes effectively to other permutation-invariant MIL architectures beyond ABMIL.}

\newpage

\section*{Discussion}

In this study, we introduce HIPPO, an explainable AI method designed to enhance the interpretability and trustworthiness of weakly supervised multiple instance learning models in computational pathology. Our results demonstrate HIPPO's ability to uncover hidden biases, quantify the impact of specific tissue regions on model predictions, and bridge the gap between computational outputs and clinically relevant insights. HIPPO offers a generalizable framework for interpreting, auditing, and stress-testing MIL models, enabling rigorous evaluation and greater confidence in pathology AI systems.

One of the key strengths of HIPPO lies in its capacity to reveal model-specific limitations that are not apparent from performance metrics or attention mechanisms alone. In our evaluation of metastasis detection models, we uncovered surprising variations in how different foundation models process histological information. For instance, some models showed a strong reliance on peritumoral tissue, while others demonstrated unexpected insensitivity to small tumor regions. These findings underscore the importance of rigorous model evaluation beyond standard performance metrics and highlight potential pitfalls in clinical deployment.

The revelation that high-attention regions can sometimes have counterintuitive effects on prognostic predictions is particularly striking. This disconnect between attention and model output challenges the common practice of using attention maps as a primary means of model interpretation. Our results suggest that regulatory bodies and clinical teams should exercise caution when relying solely on attention-based explanations and should consider incorporating quantitative impact assessments, such as those provided by HIPPO, in their evaluation processes. For example, one may use \textit{HIPPO-knowledge} to quantify the effect of high attention regions on model predictions.

HIPPO's ability to verify that models have learned biologically relevant information, as demonstrated by our analysis of TILs in prognostic models, is crucial for building trust in AI-driven clinical tools. This alignment between model behavior and established biological knowledge provides a foundation for explaining model decisions to clinicians and patients, potentially facilitating the integration of AI tools into clinical workflows. It is also possible to use HIPPO's de novo search to identify sets of patches from which expert pathologists could interpret manually to identify new tissue biomarkers.

The application of HIPPO to simulate the effects of autologous TIL therapy in melanoma patients showcases the potential for virtual experimentation in computational pathology. As foundation models and ABMIL methods improve, this approach could have far-reaching implications for personalized medicine, offering a computational method to predict treatment responses and guide therapy selection. However, it is important to note that these simulations are based on model predictions and would require extensive clinical validation before they can be considered for patient care.

Our findings suggest several key considerations for the future development and deployment of ABMIL models in clinical settings: (1) Model developers should incorporate robustness to tissue heterogeneity and small tumor regions as explicit design goals, potentially through targeted data augmentation using HIPPO-based counterfactuals; (2) Regulatory approval processes for AI tools in pathology may consider including comprehensive evaluations of model behavior across diverse tissue contexts, going beyond aggregate performance metrics; (3) The implementation of AI tools in clinical practice should be accompanied by clear explanations of model strengths and limitations, with HIPPO-like analyses providing quantitative assessments of model reliability for specific tissue types or patient subgroups; (4) Post-deployment monitoring of AI models should include ongoing analysis of model behavior in real-world settings, with HIPPO offering a means to detect potential shifts in model performance or the emergence of unexpected biases.

While our study demonstrates the potential of HIPPO, several limitations must be acknowledged. First, the counterfactual scenarios generated by HIPPO, while informative, may not always reflect biologically plausible tissue alterations. Future work should focus on refining these interventions to more closely mimic realistic tissue changes. Second, HIPPO operates at the resolution of patches, which limits its ability to capture finer-grained tissue-level details. \rebuttal{Third, although we demonstrate HIPPO's applicability beyond ABMIL using a permutation-invariant ViT model, HIPPO currently depends on architectures that operate on patch-level embeddings and do not use spatial information. Extending HIPPO to end-to-end or non-MIL models, including foundation models with spatial reasoning or slide-level supervision, remains an important direction for future research. }

\rebuttal{Additionally, our analysis was limited to a specific set of foundation models and datasets. Broader evaluation across diverse pathology tasks and model architectures is needed to fully characterize HIPPO’s capability in real-world settings as well as the generalizability of our findings. Importantly, HIPPO’s interpretive power is ultimately constrained by the representational capacity of the underlying AI model. Because HIPPO generates counterfactual predictions using the model’s learned associations, any limitations in the model’s ability to capture complex biological relationships will naturally be reflected in its explanations. As a result, HIPPO is best viewed as a rigorous framework for understanding model behavior and generating testable hypotheses, rather than a standalone tool for definitive biomarker discovery in settings where the underlying model's predictive power is not yet robustly established.}

Looking ahead, this work opens several promising directions for future research. Integrating HIPPO with multimodal data, such as genomic and clinical information, could deepen its capacity to reveal insights into both model behavior and biological relevance. While HIPPO identifies important patches de novo, interpreting these patches often requires the expertise of trained pathologists to discern the shared features driving model predictions. Alternatively, observational statistics derived from the patches, such as nuclear pleomorphism from nuclear segmentation or tissue type classifications, could help identify key features within HIPPO-identified patches. Incorporating matched spatial transcriptomics data offers another compelling avenue \autocite{jaume2024hest}, potentially enabling biomarker discovery by linking sub-visual molecular features with morphological characteristics. Such integrations could help overcome the patch-level resolution limit of HIPPO, offering more granular insights into the interplay between molecular features and patient prognosis.

Another critical avenue involves leveraging HIPPO as a powerful auditing tool for ABMIL models. By systematically identifying model limitations, HIPPO can uncover hidden biases, weaknesses in feature representation, or instances where predictions rely on irrelevant or spurious features. This level of scrutiny can guide targeted refinements, such as fine-tuning models to correct specific deficiencies or retraining with more representative data. Moreover, HIPPO’s ability to highlight important patches provides a transparent framework for validating whether the model aligns with clinical expectations and pathology expertise. These improvements have the potential to not only enhance model robustness but also foster trust and reliability in clinical applications, paving the way for safer and more interpretable deployment in real-world settings.

In conclusion, HIPPO represents a major advance in the ability to interpret AI models in computational pathology. By providing a quantitative framework for assessing the impact of specific tissue regions on model predictions, HIPPO offers a powerful tool for uncovering model limitations, verifying biological relevance, and biomarker discovery for myriad clinical applications. As the field of computational pathology continues to evolve, quantitative methods like HIPPO will be crucial in ensuring that AI tools are deployed responsibly and effectively in healthcare settings.


\phantomsection
\section*{Methods}
\setcurrentname{Methods} 
\label{sec:methods}

\subsection*{HIPPO toolkit}

HIPPO (Histopathology Interventions of Patches for Predictive Outcomes) is an explainable AI toolkit for \rebuttal{multiple instance learning (MIL) models in computational pathology, including attention-based multiple instance learning (ABMIL)\autocite{ilse2018abmil} and VisionTransformer models \autocite{el2025stamp}.} In computational histopathology, tissue from a whole slide image (WSI) is divided into small patches, which are embedded using a pre-trained model. These patch embeddings serve as input to the MIL model, which learns to map sets of patch embeddings to specimen-level labels. HIPPO is made possible by two key features of MIL: (1) models are invariant to patch order, and (2) models accommodate a variable number of patches. Taking advantage of these features of MIL, HIPPO creates counterfactual examples by adding or removing patches, allowing for the evaluation of hypothetical scenarios and the quantification of the effects of tissue region on model predictions. In other words, HIPPO simulates counterfactual WSIs by subsampling patches. 

\rebuttal{Within the HIPPO framework, users can flexibly select which patches to include or exclude, depending on their research question and settings. When annotations are available, one can consider the \textit{HIPPO-knowledge} approach}, enabling hypothesis-driven experiments. 
Patches corresponding to specific annotated regions can be removed or added to test their necessity or sufficiency for model predictions. For instance, in the present study, we utilized expert annotations of tumor regions to investigate the necessity and sufficiency of tumor presence in predicting metastasis. By removing tumor patches, we assessed their impact on model predictions to determine their criticality. Conversely, tumor patches were added to other specimens to evaluate their sufficiency in driving positive predictions. Automated methods may also be used to generate annotations. In the present report, for example, regions positive for tumor-infiltrating lymphocytes were defined using an off-the-shelf cell classification network, and these regions were used to test hypotheses in prognostic models.

In many cases, annotations are unavailable. In these cases, attention-based methods can be used to select patches for experimentation. Patches with the highest attention scores, such as the top \SI{1}{\%} of patches, can be removed to examine their influence on predictions. The same patches can also be added to other specimens to evaluate their sufficiency for inducing specific predictions. This approach compensates for a limitation of attention mechanisms, which lack explicit information about the direction of effect. By selectively removing high-attention patches, we quantify both the direction and magnitude of their influence on model outputs.

Additionally, the HIPPO search algorithm offers an alternative for identifying critical patches de novo. This method directly evaluates the effect of each patch on the model's predictions, accommodating potential nonlinear interactions among patches. The algorithm operates as follows:
\begin{enumerate}
    \item Begin with the full set of patches for a whole slide image (WSI) and record the initial model prediction.
    \item \rebuttal{Iteratively, for each remaining patch in the WSI, we temporarily drop one patch and measure the resulting change in the model's prediction.}
    \item \rebuttal{Identify and permanently remove the patch whose impact of removal is strongest. To do so, we first compute the effect size of each patch, which is defined as the original model output minus the model output after the patch is removed. For a specimen predicted as positive, we identify the patch with the most positive effect size (argmax).  For a specimen predicted as negative, we identify the patch with the most negative effect size (argmin).}
    \item \rebuttal{Repeat the process until all patches are exhausted or a stopping criterion is met (e.g., a certain number or percentage of patches removed)}.
\end{enumerate}
The HIPPO search algorithm provides a systematic framework for identifying patches that are most critical to model predictions. This can be particularly useful when hypothesis-driven experiments reveal discrepancies between expected and actual model behavior, or when the important tissue features are not known beforehand. By identifying the regions the model prioritizes from scratch, this approach enables comparisons with domain knowledge, helps uncover unexpected decision-making patterns, and may aid in biomarker discovery. A limitation of the HIPPO search strategy is potentially long runtime, depending on the number of patches in a specimen. To address this, a batched approach can be used, wherein multiple patches are removed at each iteration. By adjusting the batch size, a trade-off can be achieved between computational efficiency and the granularity of the results. \rebuttal{Detailed algorithmic descriptions with pseudo-code for all HIPPO usages (\textit{HIPPO-knowledge}, \textit{HIPPO-attention}, \textit{HIPPO-search-positive-effect}, and \textit{HIPPO-search-negative-effect}) are provided in the \nameref{sec:supplementary_methods}.}

\subsection*{Deep neural network development}

We used attention-based multiple instance learning (ABMIL) to learn specimen-level labels from wholes slide images. For metastasis detection, we evaluated five different patch encoders: UNI \autocite{chen2024uni}, REMEDIS \autocite{azizi2022remedis}, CTransPath \autocite{wang2022ctranspath}, Phikon \autocite{filiot2023phikon}, RetCCL \autocite{wang2023retccl}, \rebuttal{and Virchow2 \autocite{vorontsov2024virchow,zimmermann2024virchow2}.} These encoders were used to embed non-overlapping $128 \times 128$ \SI{}{\micro\meter} patches, with all encoders utilizing identical patches. We standardized hyperparameters across all ABMIL models, adapting from Chen et al. \autocite{chen2024uni}. The architecture comprised a first hidden layer of 512 units and a second of 384 units, incorporating gated attention. During training, we applied a dropout rate of 0.25. The output layer performed binary classification, distinguishing between the presence and absence of metastasis. Models were trained using cross-entropy loss and the Adam optimizer with a learning rate of $1 \times 10^{-4}$, following a cosine learning rate scheduler. We used a batch size of 1 without gradient accumulation. Training continued for a maximum of 20 epochs, with the best model selected based on the highest ROC AUC on the validation set. To assess initialization variability, we trained five separate models with different random seeds for each patch encoder. For subsequent experiments, we selected the initialization yielding the highest balanced accuracy on the CAMELYON16 test set for each encoder. We visualized attention heatmaps using QuPath \autocite{bankhead2017qupath}. All models were implemented in PyTorch and trained on NVIDIA RTX 2080 Ti GPUs.

The setup for \textit{IDH} mutation classification was similar to that of metastasis classification. However, CONCH \autocite{lu2024visualconch} embeddings were used, and 10-fold Monte Carlo cross validation was employed to assess variability across models. All models were implemented in PyTorch and trained on NVIDIA RTX 2080 Ti GPUs.

For prognostic models, we used the ABMIL models defined in \autocite{chen2022porpoise}. The model was composed of a linear layer with 512 units, dropout with a rate of 0.25, and a second linear layer of 256 units. Gated attention was used. The model had four outputs, representing hazards at four points in time. Risk scores were calculated as in ref. \autocite{chen2022porpoise} and were in range $[0, 1]$, where 0 indicates lowest probability of survival. Models were all implemented in PyTorch, and training was performed on NVIDIA RTX 2080 Ti GPUs.

\subsection*{Datasets}

\subsubsection*{Breast cancer metastasis dataset}

We used the CAMELYON16 dataset \autocite{bejnordi2017camelyon16} to study breast cancer metastasis. This dataset consists of 399 images and has fine-grained tumor annotations made by expert pathologists. \rebuttal{These expert-annotated tumor segmentations provide ground-truth labels that enable systematic validation of HIPPO's causal attributions against known biological features. Most of our HIPPO experiments rely on these expert-annotated ground-truth tumor regions to validate the necessity and sufficiency of tissue regions for model predictions.} The training set was split into 90\% training and 10\% validation, stratified by the label of the specimen (i.e., normal or tumor). 
Training set consisted of 143 negative and 100 positive WSIs (52 macrometastases and 48 micrometastases).
The validation set consisted of 16 negative and 11 positive WSIS (6 macrometastases and 5 micrometastases).
We used the pre-defined test set, which consisted of 80 negative and 49 positive WSIs (22 macrometastases and 27 micrometastases).
In the entire dataset, there were 160 metastasis-positive specimens. There was an average tumor area of \SI{12.26}{\milli\meter\squared} (std. dev. \SI{34.04}{\milli\meter\squared}; minimum \SI{0.008}{\milli\meter\squared}; and maximum \SI{276.09}{\milli\meter\squared}). All 399 slides had pixel spacings between \SI{0.226} and \SI{0.243}{\frac{\micro\meter}{px}} (MPP). The WSIs had $10,250 \pm 6,672$ patches (mean $\pm$ standard deviation), where each patch was $128 \times 128$ \SI{}{\micro\meter}.

\subsubsection*{Prognostic datasets}

Prognostic models were trained and evaluated using the invasive breast carcinoma (BRCA) and cutaneous melanoma (SKCM) studies from The Cancer Genome Atlas. In TCGA BRCA, 1,022 WSIs from 956 patients were used (130 death events), and in TCGA SKCM, 268 slides from 230 patients were used (89 death events). Overall survival time and censoring was used and retrieved from the code repository\footnote{\url{https://github.com/mahmoodlab/PORPOISE}} of ref. \autocite{chen2022porpoise}. The training and validation splits for cross validation were accessed from the same code repository. The WSIs in TCGA BRCA had $11,260 \pm 6,544$ patches (mean $\pm$ standard deviation). The WSIs in TCGA SKCM had $14,153 \pm 7,471$ patches.

\subsubsection*{\textit{IDH} mutation dataset}

The EBRAINS dataset \autocite{roetzer2022digital}, comprising 795 patients and 873 specimens with known \textit{IDH} mutation status, was used to develop \textit{IDH} mutation classification models. The dataset included 540 \textit{IDH} wildtype and 333 \textit{IDH} mutant specimens, spanning 508 glioblastomas, 189 astrocytomas, and 176 oligodendrogliomas. All WSIs were in NDPI format with physical spacing of \SI{0.227}{\micro\meter}. Specimens were partitioned for 10-fold Monte Carlo cross-validation using the CLAM toolkit \autocite{lu2021clam}. The WSIs had $9,100 \pm 6,878$ patches (mean $\pm$ standard deviation), where each patch was $128 \times 128$ \SI{}{\micro\meter}.

\subsection*{Whole slide image processing}

Whole slide images were read using OpenSlide \autocite{goode2013openslide}, and a modified version of the CLAM toolkit \autocite{lu2021clam} was used to segment tissue and calculate patch coordinates. Regions of tissue are identified to not spend computational resources on glass regions of the WSI. The image is converted to HSV color model (hue, saturation, value/brightness). The saturation channel is smoothed and thresholded to create a binary tissue image. Non-overlapping patches coordinates of $128 \times 128$ \SI{}{\micro\meter} were calculated within the tissue regions. The CLAM toolkit \autocite{lu2021clam} was modified to create patches at uniform physical sizes. The size of a patch in pixels can vary based on the spacing ($\SI{}{\micro\meter}/\text{px}$, MPP), and the patch size in base pixels is calculated using,
\begin{equation}
    \text{Patch size}\thinspace(\text{px}) = \frac{\text{Patch size}\thinspace(\SI{}{\micro\meter})}{\text{WSI spacing}\thinspace(\frac{\SI{}{\micro\meter}}{\text{px}})}
\end{equation}
The $128 \times 128$ \SI{}{\micro\meter} patches from CAMELYON16 were then embedded using \rebuttal{six} pre-trained models (embedding dimensions in parentheses): UNI (1024) \autocite{chen2024uni}, REMEDIS (4096) \autocite{azizi2022remedis}, Phikon (768) \autocite{filiot2023phikon}, CTransPath (768) \autocite{wang2022ctranspath}, RetCCL (2048) \autocite{wang2023retccl}, and and Virchow2 (1280) \autocite{vorontsov2024virchow,zimmermann2024virchow2}. The TCGA-BRCA and TCGA-SKCM specimens for prognostic models were embedded using the UNI model, and the glioma specimens from EBRAINS were embedded using the CONCH model. When embedding a slide, patches were read directly from the WSI file. A batch of 64 patches was read from the WSI, and then that batch was processed by the model to compute embeddings. The embeddings of all patches in a WSI were concatenated into one array and saved to disk for reuse. These embeddings served as inputs to all ABMIL models in the present report.

\subsection*{HIPPO experiment details}

\subsubsection*{Testing the necessity of tumor regions}

The degree to which tumor regions influence ABMIL models for metastasis detection remains unclear. To test the necessity of tumor regions, all of the tumor in the 49 tumor-positive specimens was removed, and changes in model outputs were recorded. The embeddings of all patches intersecting with expert tumor annotations were removed. See Supplementary Fig. \ref{fig:supp_hist_patches_with_mets} for the histogram of the number of patches removed from each WSI. Once all of the tumor patches were removed from the bag of embeddings, the specimen was called ``negative'' for metastasis. The modified bags of embeddings were run through the model, and outputs were recorded. The true negative rate (specificity) was calculated as the ratio of true negative detections to all negative samples. In this case, as all samples were negative, the true negative rate was the proportion of specimens called negative by the model. This was done for all patch embedding tested in the present report.

\subsubsection*{Testing the sufficiency of tumor regions}

The sufficiency of tumor regions for metastasis detection remains unclear. We evaluated this in two ways: by evaluating the use of only tumor tissue from positive specimens (n=49), and by embedding metastatic patches from positive specimens (n=49) into negative specimens (n=80). In the first method, we removed all patches that did not intersect with the expert tumor annotations. This evaluated the hypothetical scenario that the specimen contained tumor and no other type of tissue. The labels of all specimens remained ``positive'', and model outputs were recorded. Sensitivity was measured as the proportion of positive model predictions.

In the second method, we created counterfactual examples of metastasis-positive specimens (n=3920) from normal specimens (n=80). All combinations were evaluated: the patches that intersected expert tumor annotations from each positive slide were added to each negative slide, making a total of 3920 counterfactual examples (80 negative $\times$ 49 positive specimens). Each of these counterfactual examples was labeled ``positive'' because they contained tumor. These counterfactual examples were then run through the ABMIL models, and outputs were recorded. Sensitivity was measured as the proportion of positive model outputs.

\subsubsection*{Testing the effect of tumor size}

The extent to which tumor size affects specimen-level metastasis detection is incompletely understood. Conventionally, this analysis is limited to existing specimens. We explore a richer set of tumor sizes using counterfactual examples. First, we evaluated the effect of a single $128 \times 128$ \SI{}{\micro\meter} tumor region in normal and metastatic specimens. The tumor region was taken from specimen \texttt{test\textunderscore001} at the coordinates (37878, 63530, 38444, 64096), indicating minimum X, minimum Y, maximum X, and maximum Y.  For normal specimens, we added the embedding of this one patch into each of the 80 normal specimens and fed these bags of embeddings to the ABMIL model. Sensitivity was measured as the proportion of positive model predictions. We also evaluated this in the context of positive specimens. First, all tumor patches intersecting with expert tumor annotations were removed, and the single patch embedding was added to the bags of embeddings. 48 positive samples were used -- the specimen that the patch came from was not included. Sensitivity was measured as the proportion of positive predictions.

In addition, the effect of each individual tumor patch was evaluated for metastasis detection. In the positive slides (n=49), all tumor patches intersecting expert tumor annotations were removed to render the slide negative for metastasis. Then, each tumor patch that was fully contained by the tumor annotations was added to the bag of embeddings one at a time, and model outputs were recorded. Model probabilities for tumor were recorded.

Lastly, the size of tumor was evaluated by sampling increasing numbers of tumor patches. First, all tumor patches intersecting with expert tumor annotations were removed. Then, tumor patches fully contained by the annotations were randomly sampled and added back to the bag of embeddings. This was evaluated over multiple numbers of sampled patches (i.e., 1, 2, 4, 8, 16, 32, 64). Sensitivity was evaluated as the proportion of positive predictions.

\subsubsection*{Identifying the largest unseen tumor}

Motivated by the graded effect of tumor size on metastasis detection performance, we sought to identify the largest area of tumor that would still result in a negative prediction by the ABMIL models. To accomplish this, we used a HIPPO search algorithm. First, all patches that intersected the expert tumor annotation were removed, to render the specimen ``negative'' for metastasis. Then, tumor patches were added to the specimen one at a time, and model outputs were assessed. The tumor patch that resulted in the lowest model probability of tumor was kept in the bag, and the next round of the search was initiated. This was repeated until the model probability of tumor was greater than 0.5, which would trigger a positive prediction. The set of tumor patches that were in the bag prior to reaching a threshold of 0.5 were considered the largest area of tumor that could be present while maintaining a negative predictions.

\subsubsection*{Testing the effect of adipose tissue on metastasis detection}

Upon inspection of attention maps for the CTransPath-based metastasis detection model, adipose regions had high attention in a false negative, leading us to hypothesize that adipose regions were driving the false negative in that specimen. Attention alone could not allow us to address this hypothesis, but HIPPO could. The adipose regions were annotated in QuPath \rebuttal{and verified by two board-certified histopathologists}. Patches that intersected with the adipose region were removed, while ensuring that no tumor patches were removed. To measure the effect of this adipose tissue in other specimens, patches intersecting with the adipose annotation were added to the other 48 metastasis-positive slides, and the number of changes from true positive to false negative were recorded.

\subsubsection*{Diagnosing shortcut learning}

We sought to evaluate how HIPPO can uncover shortcut learning and how it compares to attention in this regard. To do this, we modified the normal specimens in the CAMELYON16 dataset to include a blue square (hexadecimal color code \verb|#284283|). This is meant to mimic a plausible real-world scenario in which a pathologist marked certain slides with a blue pen. In practice, we embedded one blue square of $128 \times 128$ \SI{}{\micro\meter} using the UNI model \autocite{chen2024uni} and replicated that embedding 36 times to create a $768 \times 768$ \SI{}{\micro\meter} blue region. The embeddings of this blue region were concatenated with the patch embeddings of normal specimens. The specimens with metastasis were not modified. We reasoned that the ABMIL model would learn to distinguish normal from metastatic specimens by the presence of a blue region. To assess whether tumor regions were affecting model predictions in positive specimens, we removed all patches intersecting with tumor annotations in positive specimens and recorded model outputs. To visualize attention maps, we saved patch-wise attention weights in GeoJSON format and visualized the maps in QuPath \autocite{bankhead2017qupath}. We also used the search strategy \textit{HIPPO-search-positive-effect} to identify the regions with the highest effect sizes \textit{de novo}. We also did this using a UNI-based ABMIL model trained on the original, unaltered CAMELYON16 dataset, trained using the same hyperparameters and random seed.

\subsubsection*{Identifying prognostic regions and comparing with attention}

We sought to compare the effectiveness of attention and HIPPO for identifying tissue regions related to predicted prognosis. TCGA BRCA and SKCM data were used in these experiments. For attention, regions assigned the top 1\% of attention scores were selected. For HIPPO, the search strategy \textit{HIPPO-search-positive-effect} was used to identify the regions most contributing to high risk in high-risk specimens, and the search strategy \textit{HIPPO-search-negative-effect} was used to identify the regions most contributing to low risk in low-risk specimens. Low and high risk were defined as the first and fourth quartiles of predicted risk scores, respectively. The first 1\% of patches identified by the HIPPO search algorithms were selected for evaluation. To quantify the effect of the selected regions on predicted prognosis, we calculated the difference between the predicted prognosis on the original specimens and the predicted prognosis on the specimens with the selected regions removed.

\begin{equation}
    \text{Risk contribution of ROI} = \text{Risk using original WSI} - \text{Risk when ROI is removed}
\end{equation}

Positive values indicated that the regions contributed to higher risk, and negative values indicated that the regions contributed to lower risk. Independent two-sided t-tests were used to assess the significance of differences between attention and HIPPO.

\subsubsection*{Effect of TILs on prognostic models}

In prognostic models, we measured the effects of tumor-infiltrating lymphocytes (TILs) on model behavior. The number of TILs was quantified using the same approach as Ref. \autocite{chen2022porpoise}. Briefly, HoVer-Net \autocite{graham2019hovernet} was used to outline and label the nuclei in TCGA BRCA and SKCM WSIs. The model labels nuclei as one of six categories: tumor epithelium, lymphocyte, stroma, necrosis, normal epithelium, and unknown. Each $128 \times 128$ \SI{}{\micro\meter} was called TIL-positive if it contained more than 20 cells, more than 10 immune cells, and more than 5 tumor cells. In TCGA BRCA, HoVer-Net failed for 12 WSIs, some of which were missing pixel spacing information.

We measured the effect of TIL patches on predicted prognosis in TCGA BRCA and SKCM by either removing TILs from low-risk specimens or adding TILs to high-risk specimens, where low-risk was defined as samples in the first quartile of predicted risk and high-risk were samples in the fourth quartile of predicted risk. The predicted prognoses were compared before and after the intervention. To evaluate the sufficiency of TILs for predicting low risk, we added TIL patches from low-risk specimens to high-risk specimens. Risk predictions of the model were recorded, and differences were tested using paired two-sided t-tests. To assess the necessity of TIL regions, we removed TIL-positive patches from low-risk specimens and measured risk predictions. Differences were tested using paired two-sided t-tests.

\subsubsection*{Evaluating autologous TILs}

Autologous TIL therapy is a promising immunotherapy. We explored how HIPPO could be used for hypothesis generation in the context of autologous TILs in high-risk SKCM specimens (n=67). We sought to assess the degree to which prognostic ABMIL models are affected by the number of TILs in a specimen. We do not claim to assess the efficacy of autologous TILs through HIPPO. The embeddings of TIL-positive regions were replicated $2\times$, $10 \times$, $20 \times$, or $100 \times$, and the change in predicted risk was measured:

\begin{equation}
    \text{Change in Risk} = \text{Risk with autologous TILs} - \text{Risk with original WSI}
\end{equation}

Negative values indicated that the addition of TILs decreased risk. The change in risk from baseline was assessed using paired two-sided t-tests.

\subsubsection*{Explaining misclassifications in \textit{IDH} mutation classification}

We analyzed histopathology specimens with known \textit{IDH} mutations but misclassified as negative by the models (i.e., false negative, FN).
The objective was to evaluate whether removing the top \SI{20}{\%} of patches, identified by attention or HIPPO search, could increase the likelihood of a correct classification. To reduce running time of experiments, we analyzed FN specimens with fewer than \SI{20000}{} patches (n=38). For each specimen, the softmax probability for \textit{IDH} mutation was recorded. Next, the \SI{20}{\%} of patches with the highest attention scores were removed, and the updated model probability was documented. In addition, the top \SI{20}{\%} of patches identified by \textit{HIPPO-search-negative-effect} were removed, and the updated model probability was recorded (search algorithm described below). The final \textit{IDH} mutation probabilities after the different methods of patch removal were compared across baseline, attention, and HIPPO conditions. Cohen’s \textit{d} was used to assess effect sizes, and independent two-sided t-tests evaluated statistical significance between conditions. The experiment was conducted exclusively on specimens from the test set of each fold during the 10-fold Monte Carlo cross-validation, ensuring that no specimens from the training or validation sets were included.

The \textit{HIPPO-search-positive-effect} algorithm iteratively removed individual patches, replacing each patch after recording its impact on mutation probability. The patch whose removal most increased the probability of \textit{IDH} mutation was permanently excluded, with this process repeated until \SI{20}{\%} of patches were removed. The search was performed using NVIDIA A100 GPUs.

\rebuttal{For applying HIPPO to transformers, we used the Vision Transformer architecture which was adapted to the permutation-invariant MIL setting by excluding the positional embeddings \autocite{el2025stamp}. To extract attention scores from the ViT model, we considered two approaches: (1) Attention (Last), which uses the raw attention weights from the CLS token to each patch token in the final self-attention layer, (2) Attention (Rollout), which is computed by recursively multiplying CLS-centric attention across all the transformer layers, following prior work\autocite{abnar2020quantifying}.}

\subsubsection*{Identifying regions driving positive \textit{IDH} mutation prediction}

We evaluated the extent to which attention and HIPPO search identified patches driving positive predictions, focusing on specimens with positive predictions in the first Monte Carlo cross-validation fold (n=25). The \textit{HIPPO-search-positive-effect} algorithm was applied, removing patches in batches of 10 per iteration to accelerate runtime. Patches were ranked by their impact on \textit{IDH} mutation probabilities, with the 10 patches whose removal most reduced the prediction probability excluded at each step. The \textit{IDH} mutation probability was recorded after each batch, continuing until \SI{90}{\%} of patches were removed. A similar approach was used for attention, removing patches in descending order of attention scores in batches of 10 and recording model outputs after each removal. Model outputs for each specimen were visualized as line plots, comparing results from HIPPO and attention. Heatmaps of attention and HIPPO outputs were viewed using QuPath for qualitative assessment.

To assess the efficiency of HIPPO in identifying critical patches, we conducted an experiment to determine how many patches needed to be removed to reduce the model probability below 0.4, hypothesizing that HIPPO would require fewer patches than attention due to its ability to better capture the most important regions driving predictions. The HIPPO and attention outputs from above were used in this experiment. The ratio of patches required (i.e., number of patches required normalized by number of patches in specimen) were compared between attention and HIPPO using an independent two-sided t-test and visualized using a scatter plot.

\section*{Data availability}
The CAMELYON16 dataset is available at \url{https://camelyon17.grand-challenge.org/Data/} under the CC0 license (public domain). The results shown here are in whole or part based upon data generated by the TCGA Research Network: \url{https://www.cancer.gov/tcga}. Clinical data and whole slide image files can be accessed at \url{https://portal.gdc.cancer.gov}. Training and validation splits for prognostic models were accessed at \url{https://github.com/mahmoodlab/PORPOISE}.

\section*{Code availability}

A Python package implementing HIPPO is available at \url{https://github.com/kaczmarj/HIPPO} and is licensed under the terms of the 3-Clause BSD License. HIPPO documentation is published at \url{https://github.com/kaczmarj/HIPPO} under the terms of the Creative Commons Attribution-NonCommercial-ShareAlike 4.0 International copyright license (CC BY-NC-SA 4.0). Model weights and inference code are available at the following repositories: UNI (\url{https://huggingface.co/MahmoodLab/UNI}), REMEDIS (\url{https://github.com/google-research/medical-ai-research-foundations}), Phikon (\url{https://huggingface.co/owkin/phikon}), CTransPath (\url{https://github.com/Xiyue-Wang/TransPath}), RetCCL (\url{https://github.com/Xiyue-Wang/RetCCL}), and Virchow2 (\url{https://huggingface.co/paige-ai/Virchow2}). The ViT model used in the \textit{IDH} mutation classification experiment is available at the following repository: \url{https://github.com/KatherLab/STAMP}. Model weights for the models trained for this report will be deposited to online repositories.

\section*{Acknowledgments}

This research was supported by National Science Foundation (NSF) grant IIS2212046, National Institutes of Health (NIH) grant UH3CA225012, R01EB035934, and Stony Brook Profund 2022 seed funding. JRK was also supported by the Medical Scientist Training Program at Stony Brook University and NIH grant T32GM008444 (NIGMS). This work was performed with assistance from CSHL Shared Resources, including the CSHL Animal \& Tissue Imaging Shared Resource, which are supported by the Cancer Center Support Grant 5P30CA045508. We would also like to acknowledge the Department of Biomedical Informatics at Stony Brook University and the Simons Center for Quantitative Biology at Cold Spring Harbor Laboratory.

\section*{Author contributions}
JRK and PKK conceived of the method. JRK, CK, SG, SL, and PKK planned the experiments. JRK, CK, and SG wrote the code and ran all experiments. All authors interpreted the results of experiments. ZZ and DS provided expert annotations for select histopathology patches. SL, JHS, and PKK supervised the project. JRK, CK, SG, SL, JHS, and PKK contributed to the writing of the manuscript. All authors provided feedback on the manuscript and contributed to the final manuscript.

\section*{Competing interests}
The authors declare the following competing interests: J.H.S. is co-founder and chief executive officer of Chilean Wool, LLC. All other authors declare no competing interests.














\newpage

\printbibliography[segment=\therefsegment]

\end{refsegment}

\newpage

\setcounter{page}{1}
\begin{refsection}
\begin{supptable}[ht]
\renewcommand\tablename{Supplementary~Table}
\centering
\begin{tabular}{llllll}
\toprule
 & Balanced Accuracy & Sensitivity & Specificity & Precision & Weighted F1 \\
Encoder &  &  &  &  &  \\
\midrule
UNI & 0.982 (0.014) & 0.976 (0.017) & 0.988 (0.022) & 0.980 (0.033) & 0.983 (0.015) \\
REMEDIS & 0.922 (0.031) & 0.861 (0.062) & 0.983 (0.014) & 0.968 (0.024) & 0.935 (0.025) \\
Phikon & 0.907 (0.083) & 0.845 (0.158) & 0.970 (0.023) & 0.943 (0.049) & 0.920 (0.070) \\
CTransPath & 0.858 (0.016) & 0.784 (0.045) & 0.933 (0.023) & 0.879 (0.034) & 0.874 (0.013) \\
RetCCL & 0.745 (0.016) & 0.567 (0.042) & 0.922 (0.049) & 0.827 (0.074) & 0.777 (0.019) \\
\rebuttal{Virchow2} & \rebuttal{0.934 (0.235)} & \rebuttal{0.869 (0.0469)} & \rebuttal{1.000 (0.000)} & \rebuttal{1.000 (0.000)} & \rebuttal{0.949 (0.018)} \\
\bottomrule
\end{tabular}
\caption{Performance of metastasis detection models. Values shown are mean (standard deviation) across five random initializations. All models used the same hyperparameters and data splits. The training, validation, and test sets consisted of 243, 27, and 129 specimens, respectively. The test set consisted of 80 negative specimens, 22 samples with macrometastases, and 27 specimens with micrometastases.}
\label{table:supp_cam16_performance}
\end{supptable}

\clearpage
\begin{supptable}[ht]
\renewcommand\tablename{Supplementary~Table}
\centering
\begin{tabular}{llllll}
\toprule
 & Balanced Accuracy & Sensitivity & Specificity & Precision & Weighted F1 \\
Encoder &  &  &  &  &  \\
\midrule
UNI & 1.000 & 1.000 & 1.000 & 1.000 & 1.000 \\
REMEDIS & 0.949 & 0.898 & 1.000 & 1.000 & 0.961 \\
Phikon & 0.955 & 0.959 & 0.950 & 0.922 & 0.954 \\
CTransPath & 0.885 & 0.857 & 0.912 & 0.857 & 0.891 \\
RetCCL & 0.769 & 0.612 & 0.925 & 0.833 & 0.799 \\
\rebuttal{Virchow2} & \rebuttal{0.959} & \rebuttal{0.918} & \rebuttal{1.000} & \rebuttal{1.000} & \rebuttal{0.969} \\
\bottomrule
\end{tabular}
\caption{Performance of the best metastasis detection models for each encoder. This table lists the performance metrics for the single best random initialization for each encoder. Performance was calculated on the CAMELYON16 test set. The models represented here were the ones used for downstream experiments in the present report.}
\label{table:supp_cam16_performance_single_models}
\end{supptable}

\clearpage
\setlength{\tabcolsep}{1pt}
\begin{supptable}[h!]
\renewcommand\tablename{Supplementary~Table}
\centering
\small
\resizebox{\textwidth}{!}{
\begin{tabular}{p{2cm}p{3cm}p{2cm}p{2cm}p{2cm}p{2cm}}
\toprule 
Slide ID & Train data condition & Mean & Median & Min & Max  \\
\midrule
test\_002 & Original & $1.3 \times 10^{-4}$ & $4.9 \times 10^{-8}$ & $ 3.7 \times 10^{-8}$ & 0.090             \\
 & Spurious &  $9.4 \times 10^{-5}$ & $5.5 \times 10^{-5}$ & $2.1 \times 10^{-5}$ & 0.020 \\
\midrule
test\_027 & Original & $1.5 \times 10^{-4}$ & 2.8 $\times 10^{-7}$ & $1.6 \times 10^{-7}$ & $0.733$             \\
 & Spurious &$ 1.1 \times 10^{-4}$ &$ 3.8 \times 10^{-5}$ & $3.5 \times 10^{-6}$ & 0.004             \\
\midrule
test\_029 & Original & $3.0 \times 10^{-4} $&$ 1.4 \times 10^{-7}$ & $9.5 \times 10^{-8}$ & 0.644             \\
 & Spurious & $2.2 \times 10^{-4}$ & $9.4 \times 10^{-5}$ &$ 3.6 \times 10^{-5}$ & 0.029             \\
 \midrule
 test\_030 & Original &$ 1.4 \times 10^{-4}$ & $1.2 \times 10^{-7}$ & $8.0 \times 10^{-8}$ & 0.080             \\
 & Spurious & $1.1 \times 10^{-4}$ & $8.3 \times 10^{-5}$ & $2.3 \times 10^{-5} $& 0.022             \\
 \midrule
 test\_051 & Original & $1.9 \times 10^{-4}$ & $7.1 \times 10^{-7}$ & $2.1 \times 10^{-7}$ & 0.008             \\
 & Spurious & $1.3 \times 10^{-4}$ & $7.7 \times 10^{-5}$ & $5.9 \times 10^{-5}$ & 0.007             \\
 \midrule
 test\_052 & Original & $1.4 \times 10^{-4}$ & $1.2 \times 10^{-7}$ & $6.6 \times 10^{-8}$ & 0.874             \\
 & Spurious & $9.6 \times 10^{-5}$ & $6.5 \times 10^{-5}$ & $5.0 \times 10^{-5}$ & 0.009             \\
\bottomrule
\end{tabular}}
\caption{\rebuttal{Comparison of effect size statistics between a model trained on original CAMELYON16 data and a model trained on data with a spurious signal. For the latter, a 768 \(\times\) 768 \(\mu\)m blue square was artificially introduced into all negative WSIs of the CAMELYON16 dataset. For the two models, we used HIPPO-search-positive-effect to identify regions that maximally drove positive tumor predictions and their effect sizes. The analysis revealed that effect sizes in the spurious model were much lower and less localized compared to the original model.}}
\label{table:supp_spurious}
\end{supptable}
\newpage

\begin{suppfigure}[htbp]
\renewcommand\figurename{Supplementary~Figure}
    \centering
    \includegraphics[width=6.5in, clip, trim=0cm 15.5cm 0cm 0cm]{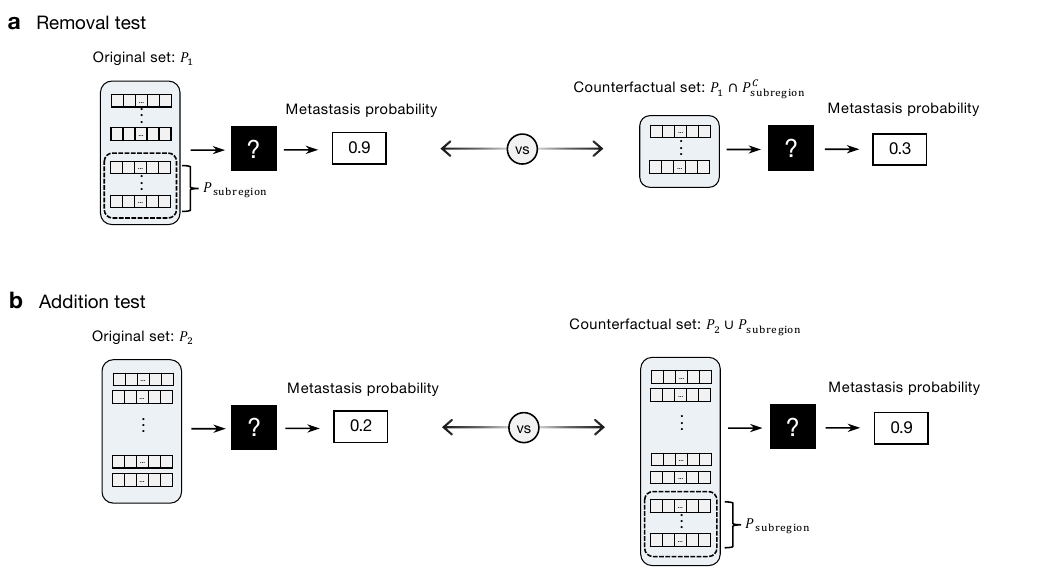}
    \\\rebuttaltemp{Added figure: clarified the two modes for patch set intervention}
    \caption{\rebuttal{Two approaches for constructing counterfactual sets in HIPPO. 
    This diagram illustrates the two primary approaches HIPPO uses to construct counterfactual inputs by manipulating patch embeddings. 
    \textbf{a,} A \textit{removal} test quantifies the influence of a subregion ($P_\text{subregion}$) by removing its patches from an original set of patch embeddings ($P_1$). A significant change in the model's output probability indicates that $P_\text{subregion}$ was highly influential for the initial prediction. 
    \textbf{b,} An \textit{addition} test assesses the sufficiency of $P_\text{subregion}$ by adding its patches (obtained from a source WSI) to a different original set (P2). A substantial increase in the model's output probability demonstrates that $P_\text{subregion}$ can drive a higher prediction.}
    }
    \label{fig:hippo-schematic-insert-delete}
\end{suppfigure}

\clearpage
\begin{suppfigure}
\renewcommand\figurename{Supplementary~Figure}
    \centering
    \includegraphics[width=0.5\textwidth]{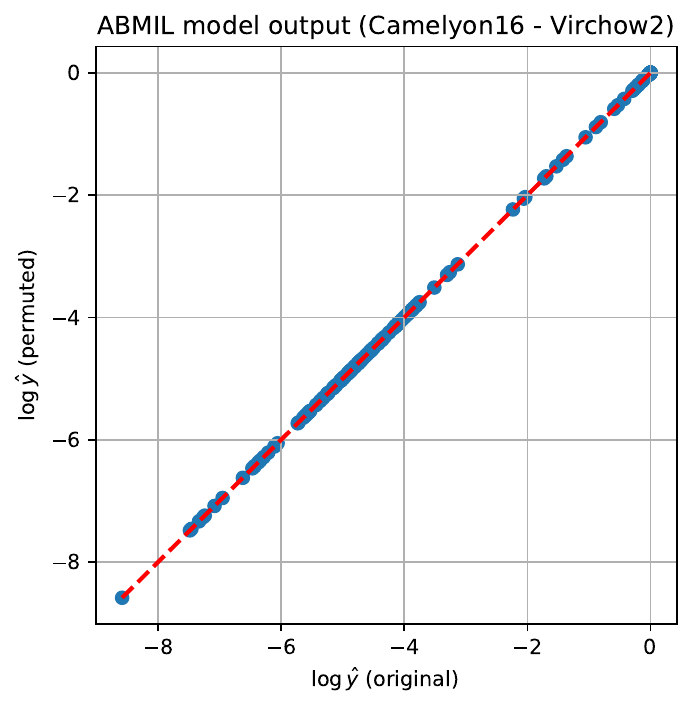}
    \\\rebuttaltemp{Added figure: Show the permutation invariant property of ABMIL model experimentally}
    \caption{\rebuttal{Effect of randomly permuting the WSI patches in the CAMELYON16 dataset on the Virchow2 ABMIL model output. Each point represents a single WSI. The model outputs before and after shuffling are the same for all slides, demonstrating that the ABMIL models are permutation invariant.}
}
    \label{fig:supp_permutation_invariance}
\end{suppfigure}

\begin{suppfigure}[htbp]
\renewcommand\figurename{Supplementary~Figure}
    \centering
    \includegraphics[width=6in]{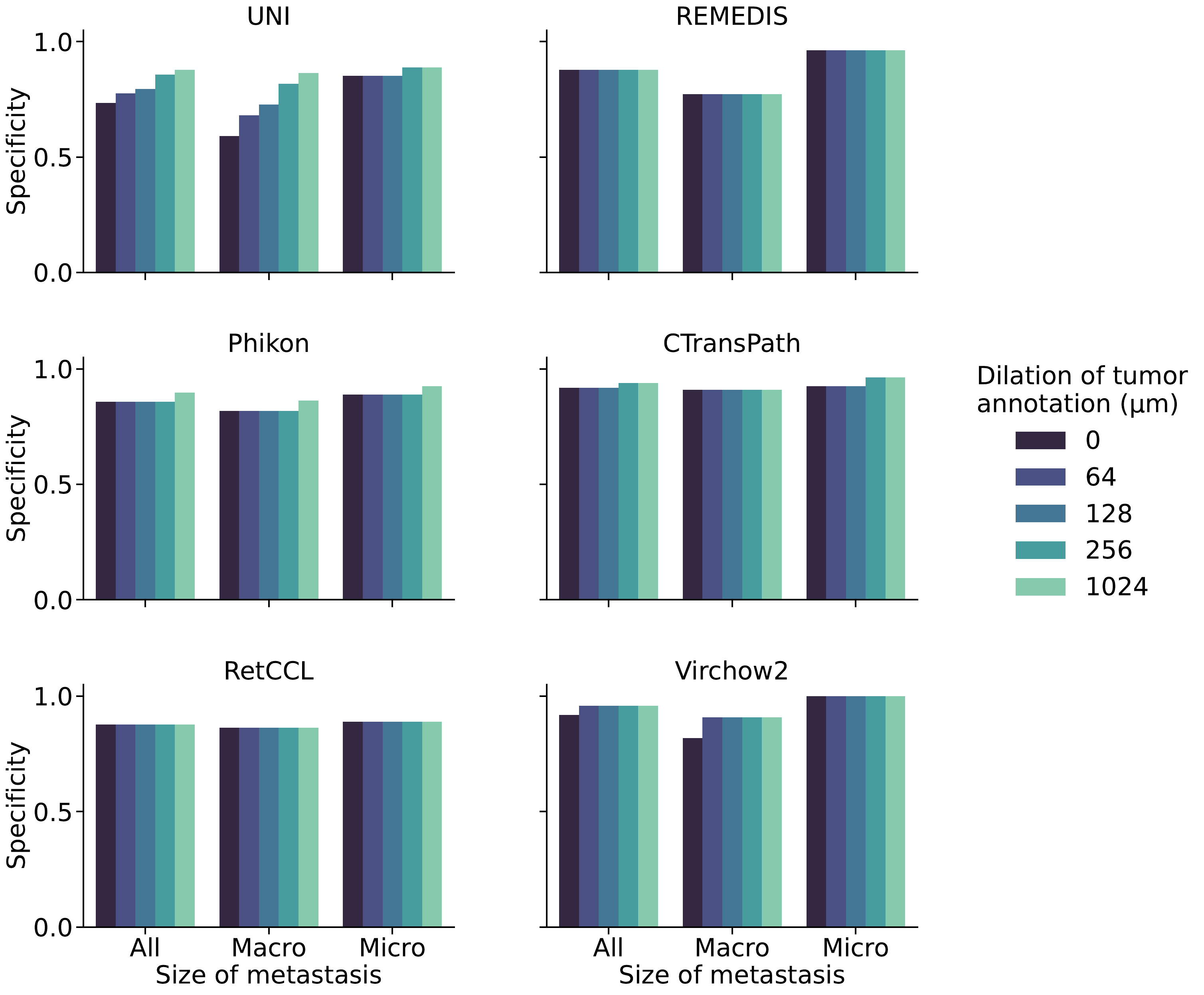}
    \\\rebuttaltemp{Modified figure: added results for Virchow2}
    \caption{Peritumoral tissue may affect metastasis detection. The expert tumor annotations in the positive specimens (n=49) of the CAMELYON16 test set were systematically dilated, and then patches that intersected with these dilated regions were removed. This effectively removed tumor tissue and varying amount of peritumoral tissue, rendering the specimens negative for metastasis. Specificity of model outputs (true negative rate) was calculated. Dilation of tumor annotations did not change model outputs in REMEDIS-based or RetCCL-based models, suggesting that peritumoral tissue did not drive model outputs. The largest dilation (i.e., \SI{1024}{\micro\meter}) increased specificity in Phikon-based and CTransPath-based models, suggesting that peritumoral was responsible to some degree for false positive predictions. The UNI-based model demonstrated a graded effect of dilation, suggesting that the tissue surrounding the tumor was driving positive model predictions. This effect was particularly strong in macrometastases (n=22). It appears that the UNI-based model relies on peritumoral tissue to some degree for positive predictions. \rebuttal{Increasing dilation from \SI{0}{\micro\meter} to \SI{64}{\micro\meter} increased specificity for the Virchow2-based model but it stayed the same after that, suggesting that peritumoral was responsible to a small degree for false positive predictions.}}
    \label{fig:supp_nec_dilations}
\end{suppfigure}

\clearpage
\begin{suppfigure}[htbp]
\renewcommand\figurename{Supplementary~Figure}
    \centering
    \includegraphics[width=1.0\textwidth]{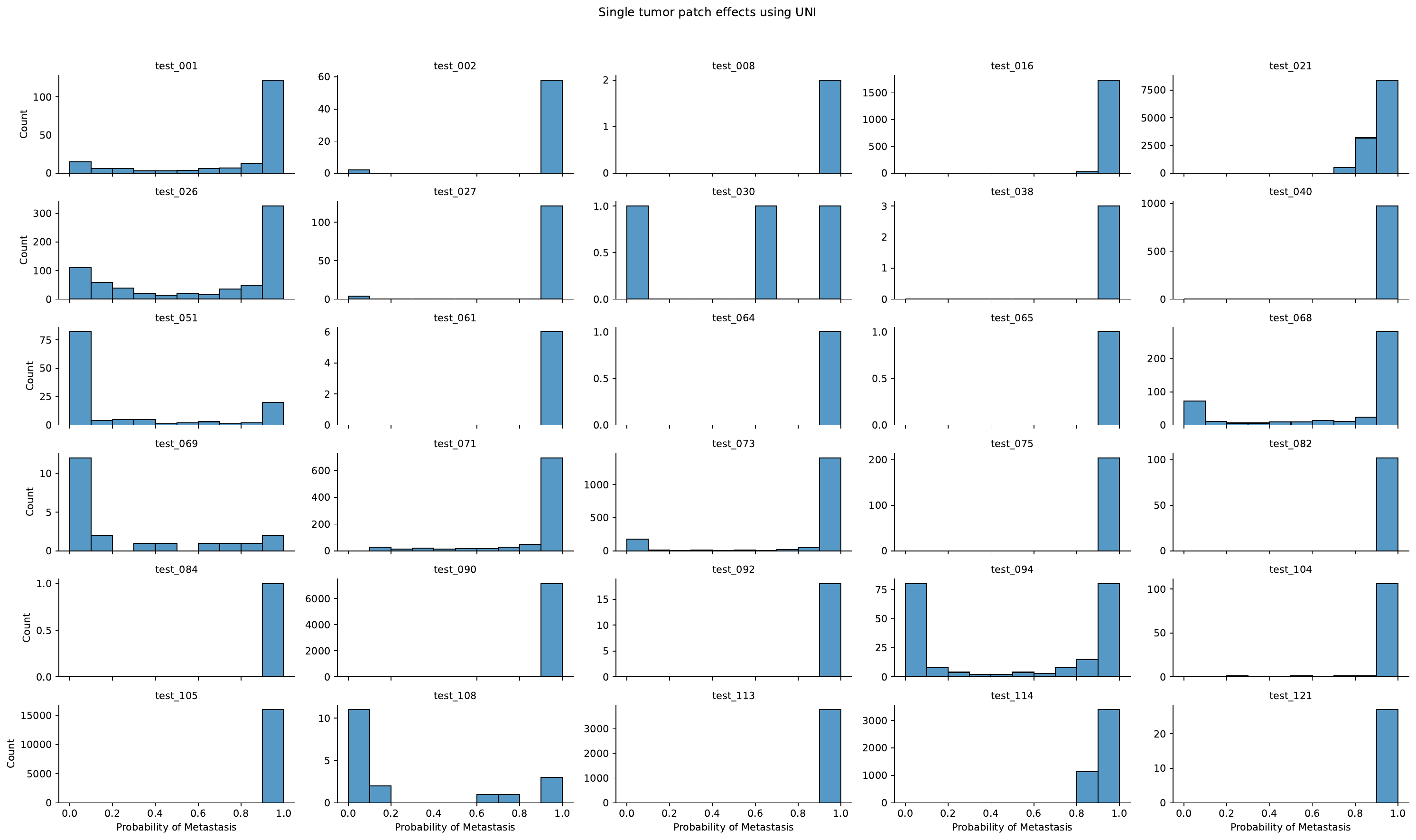}
    \caption{
    Individual tumor patch effects on metastasis detection using the UNI-based model. Histograms show model probabilities of tumor, where the $x$-axis is model probability and $y$-axis is the number of examples in the bin. Each histogram represents a different positive specimen (n=49) in the CAMELYON16 test set. First, all patches intersecting the expert tumor annotations were removed. Then, patches fully contained within the annotation were added back into the specimen one at a time, and model predictions were recorded.
    }
    \label{fig:supp_all_single_patches_uni}
\end{suppfigure}

\clearpage
\begin{suppfigure}[htbp]
\renewcommand\figurename{Supplementary~Figure}
    \centering
    \includegraphics[width=1.0\textwidth]{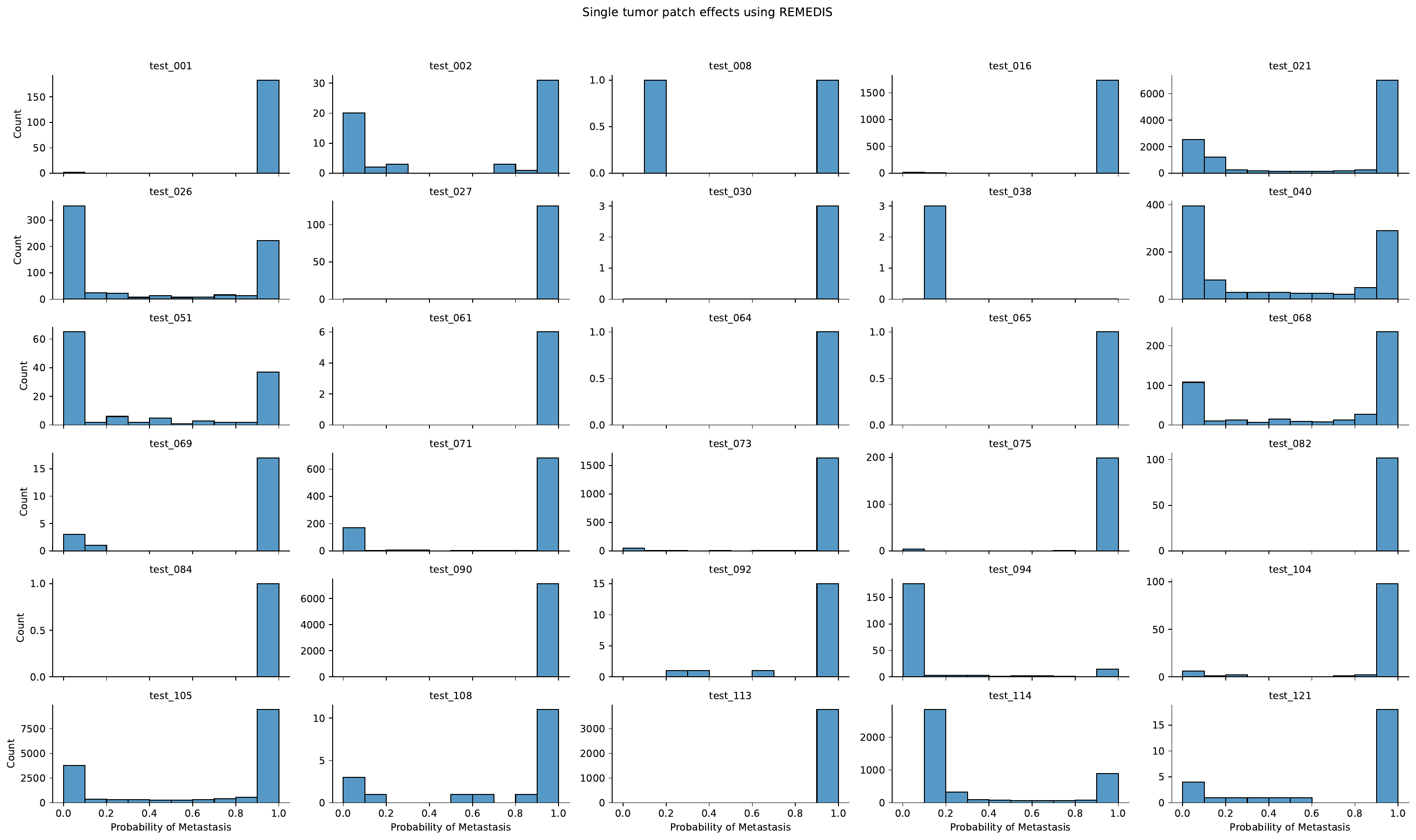}
    \caption{Individual tumor patch effects on metastasis detection using the REMEDIS-based model. Histograms show model probabilities of tumor, where the $x$-axis is model probability and $y$-axis is the number of examples in the bin. Each histogram represents a different positive specimen (n=49) in the CAMELYON16 test set. First, all patches intersecting the expert tumor annotations were removed. Then, patches fully contained within the annotation were added back into the specimen one at a time, and model predictions were recorded.
}
    \label{fig:supp_all_single_patches_remedis}
\end{suppfigure}

\clearpage
\begin{suppfigure}[htbp]
\renewcommand\figurename{Supplementary~Figure}
    \centering
    \includegraphics[width=1.0\textwidth]{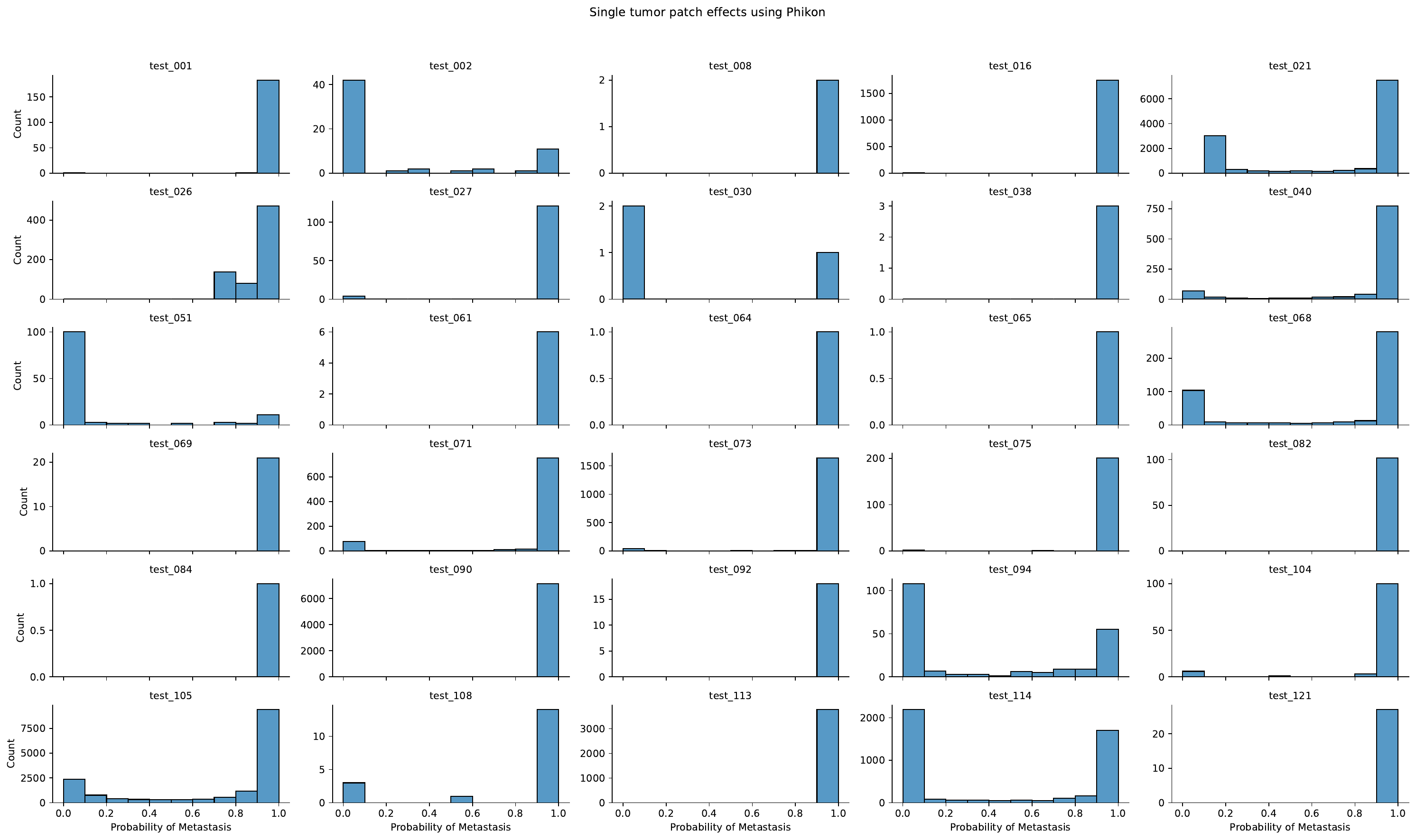}
    \caption{
    Individual tumor patch effects on metastasis detection using the Phikon-based model. Histograms show model probabilities of tumor, where the $x$-axis is model probability and $y$-axis is the number of examples in the bin. Each histogram represents a different positive specimen (n=49) in the CAMELYON16 test set. First, all patches intersecting the expert tumor annotations were removed. Then, patches fully contained within the annotation were added back into the specimen one at a time, and model predictions were recorded.
}
    \label{fig:supp_all_single_patches_phikon}
\end{suppfigure}

\clearpage
\begin{suppfigure}[htbp]
\renewcommand\figurename{Supplementary~Figure}
    \centering
    \includegraphics[width=1.0\textwidth]{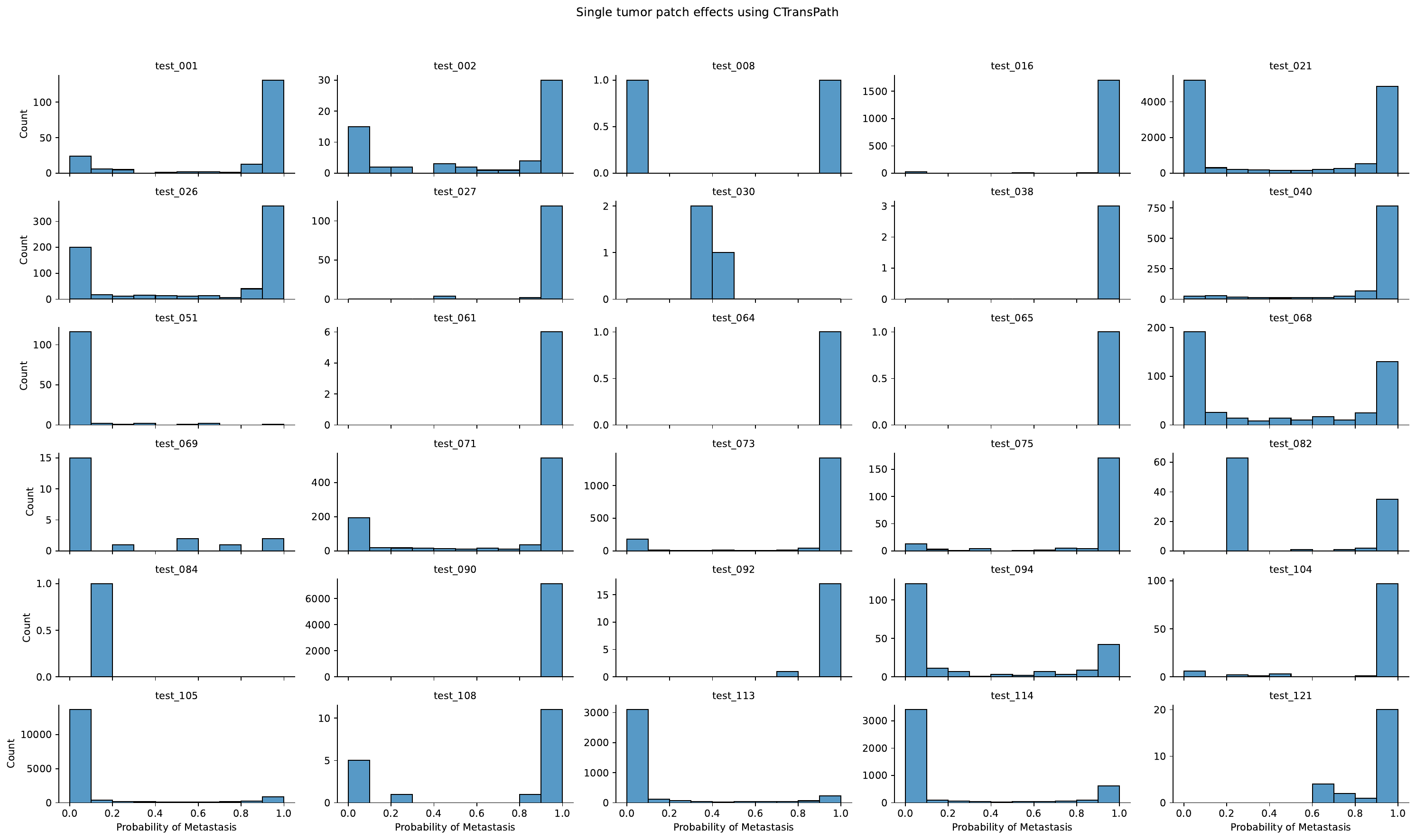}
    \caption{
    Individual tumor patch effects on metastasis detection using the CTransPath-based model. Histograms show model probabilities of tumor, where the $x$-axis is model probability and $y$-axis is the number of examples in the bin. Each histogram represents a different positive specimen (n=49) in the CAMELYON16 test set. First, all patches intersecting the expert tumor annotations were removed. Then, patches fully contained within the annotation were added back into the specimen one at a time, and model predictions were recorded.
    }
    \label{fig:supp_all_single_patches_ctranspath}
\end{suppfigure}

\clearpage
\begin{suppfigure}[htbp]
\renewcommand\figurename{Supplementary~Figure}
    \centering
    \includegraphics[width=1.0\textwidth]{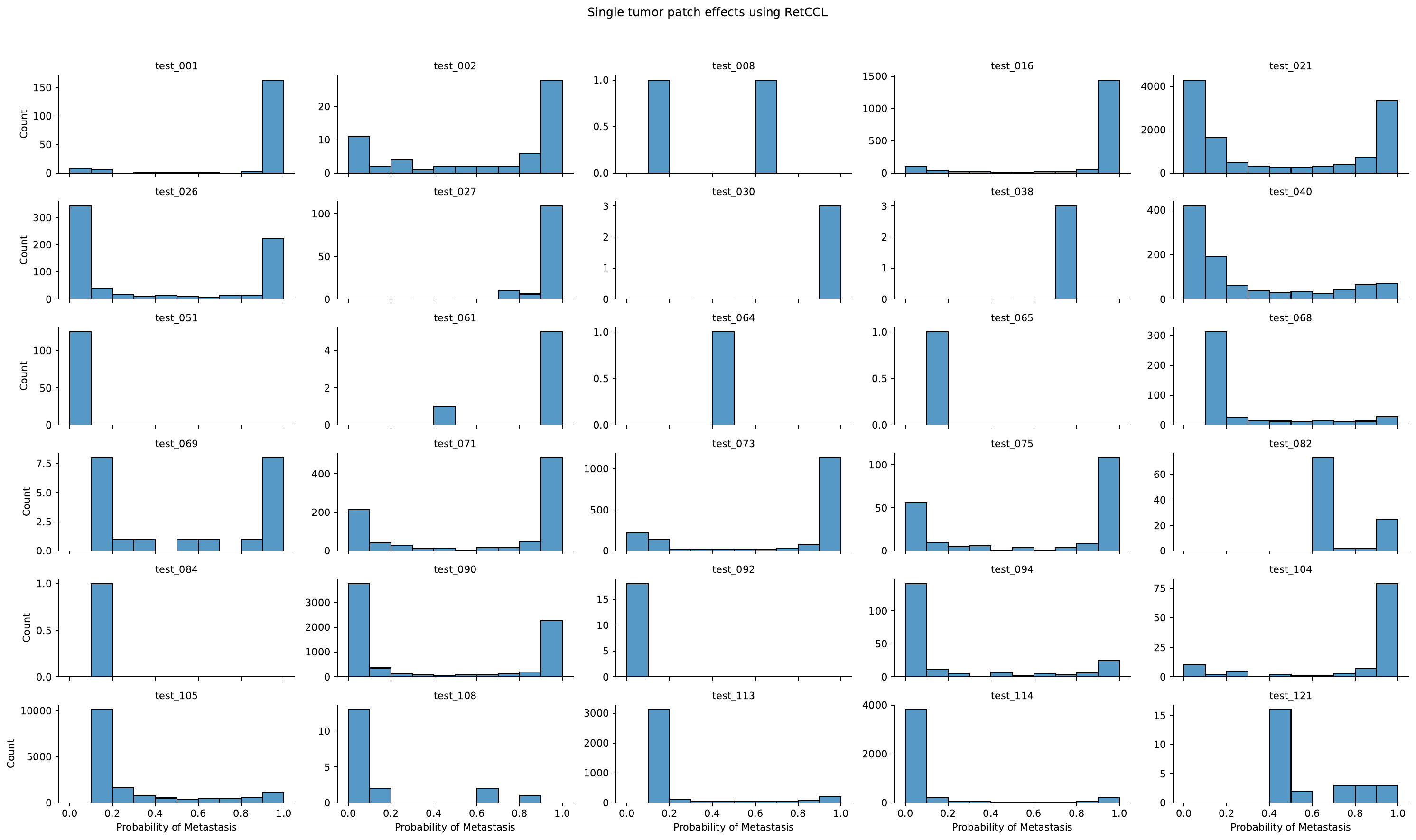}
    \caption{Individual tumor patch effects on metastasis detection using the RetCCL-based model. Histograms show model probabilities of tumor, where the $x$-axis is model probability and $y$-axis is the number of examples in the bin. Each histogram represents a different positive specimen (n=49) in the CAMELYON16 test set. First, all patches intersecting the expert tumor annotations were removed. Then, patches fully contained within the annotation were added back into the specimen one at a time, and model predictions were recorded.
}
    \label{fig:supp_all_single_patches_retccl}
\end{suppfigure}

\clearpage
\begin{suppfigure}[htbp]
\renewcommand\figurename{Supplementary~Figure}
    \centering
    \includegraphics[width=1.0\textwidth]{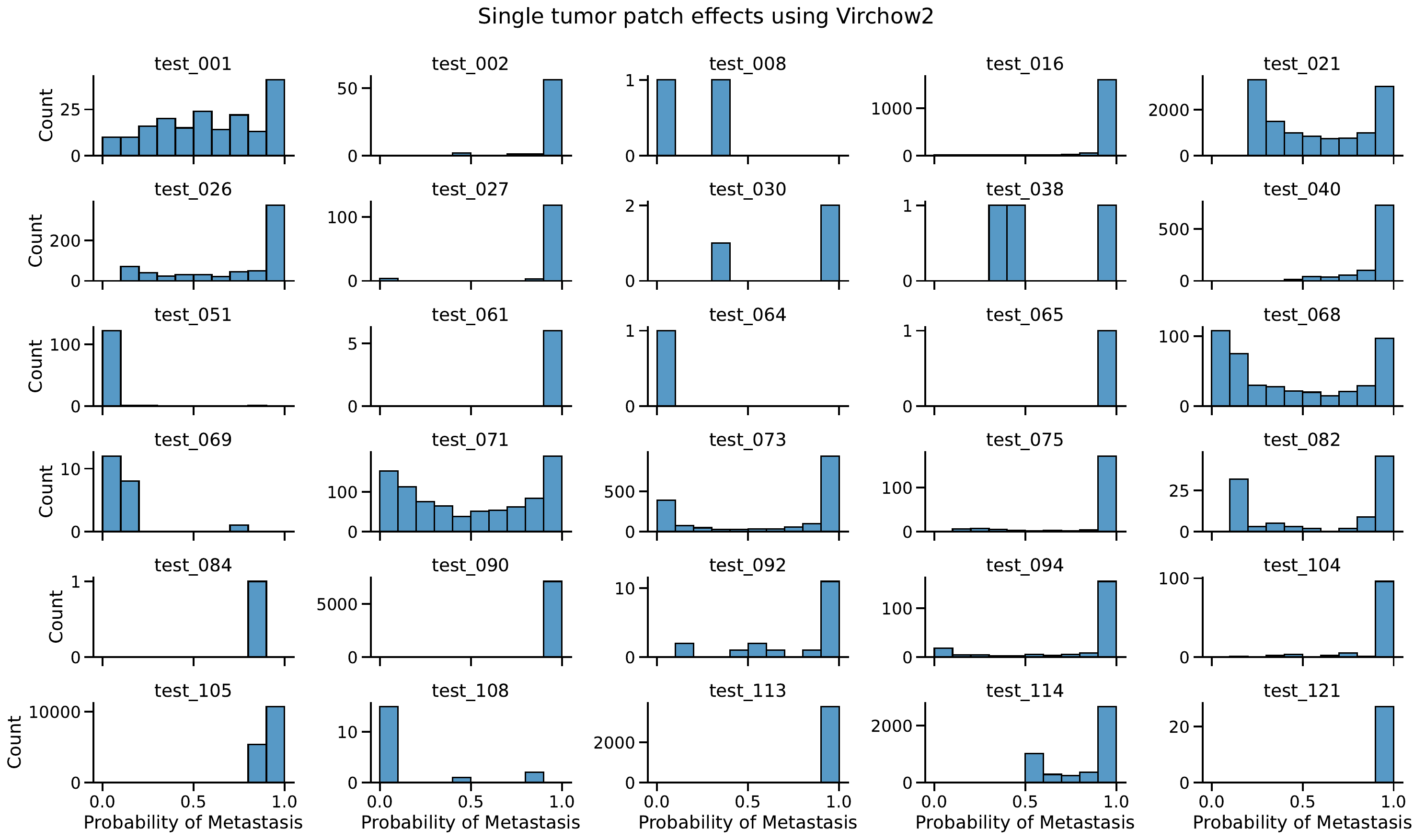}
    \\\rebuttaltemp{Added figure: show results for Virchow2}
    \caption{\rebuttal{Individual tumor patch effects on metastasis detection using the Virchow2-based model. Histograms show model probabilities of tumor, where the $x$-axis is model probability and $y$-axis is the number of examples in the bin. Each histogram represents a different positive specimen (n=49) in the CAMELYON16 test set. First, all patches intersecting the expert tumor annotations were removed. Then, patches fully contained within the annotation were added back into the specimen one at a time, and model predictions were recorded.}
}
    \label{fig:supp_all_single_patches_virchow2}
\end{suppfigure}

\clearpage
\begin{suppfigure}[htbp]
\renewcommand\figurename{Supplementary~Figure}
    \centering
    \includegraphics[width=1.0\textwidth]{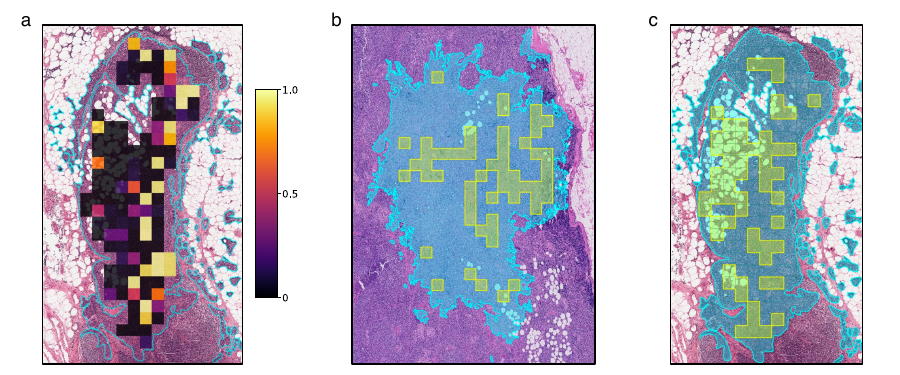}
    \caption{Tumor patches have variable effects on metastasis detection, and some tumor regions go undetected entirely. To evaluate the effect of each individual tumor patch on metastasis detection, first all patches intersecting with expert tumor annotations was removed in the positive specimens (n=49) of the CAMELYON16 test set. Then, patches that were fully contained in the tumor annotation were introduced into the specimen one at a time, and the model probability of metastasis was recorded. a, shows a representative example of model probabilities of metastasis for each tumor patch, using the UNI-based model in specimen ``test\_051''.  The expert tumor annotation is outlined in cyan. A subset of patches was sufficient to drive a positive tumor prediction (model probability $> 0.5$), but many tumor patches were insufficient to drive a positive prediction on their own. Some of these insufficient patches contained tumor epithelial cells along with adipose cells, but many did not. We also used a variant of the search algorithm \textit{HIPPO-search-negative-effect} to identify the largest set of tumor patches that can be added to a negative counterfactual while still maintaining a negative prediction. Compared to the original \textit{HIPPO-search-negative-effect} that removes patches, we progressively added tumor patches to the negative counterfactual. For each iteration, we chose a tumor patch whose effect size is lowest using the argmin function.  First, all tumor patches intersecting the tumor boundary were removed. Then, we iteratively added tumor patches back into the specimen, and kept the tumor patch that drove the lowest probability of metastasis. This continued until the model probability was greater than 0.5. These ``unseen'' tumor regions are highlighted in yellow in (b) and (c), and the tumor region is highlighted in cyan. In (b) (specimen ``test\_094''), we identified a \SI{0.95}{\milli\meter\squared} area of tumor that was undetected by the UNI-based model, and in (c), we identified a \SI{1.0}{\milli\meter\squared} area of tumor that was undetected by the UNI-based model.}
    \label{fig:supp_mets_individual_tumor_patches_unseen}
\end{suppfigure}

\clearpage
\begin{suppfigure}[htbp]
\renewcommand\figurename{Supplementary~Figure}
    \centering
    \includegraphics[width=6in]{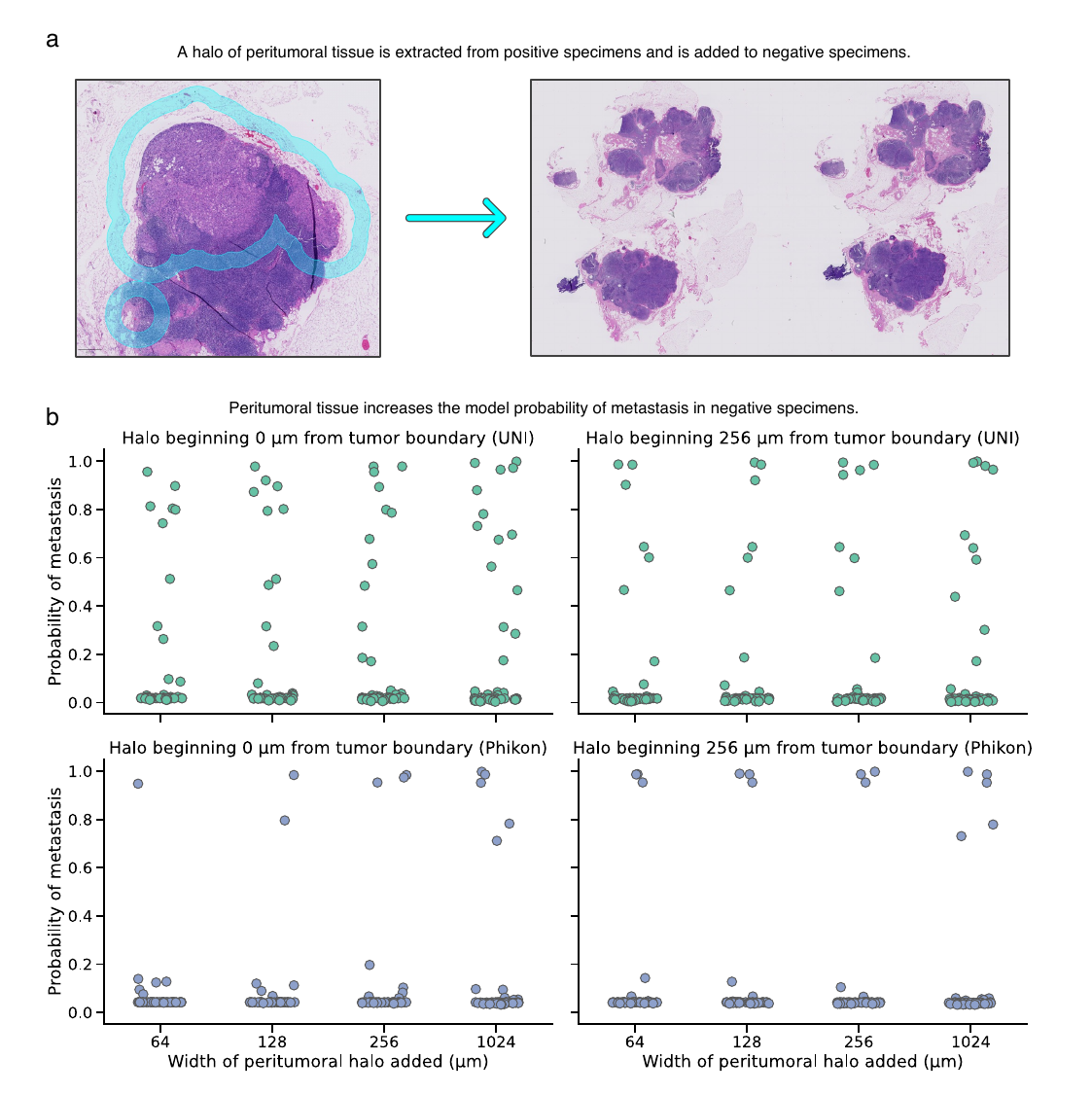}
    \caption{Non-tumor tissue is sufficient for positive detections in specimens without tumor. \textbf{a,} To evaluate the sufficiency of non-tumor tissue from positive specimens to drive positive detections in negative specimens, halos of peritumoral tissue were selected. These halos did not intersect with the expert tumor annotations, and as such were considered to be entirely non-tumor. The patches intersecting with the halo but not intersecting with tumor annotations were added to normal specimens, resulting in 3,920 counterfactual examples (80 negative $\times$ 49 positive specimens). \textbf{b,} The model's probability of metastasis was averaged across each negative specimen to evaluate the global effect of the peritumoral halo on model outputs. Four widths of halos were evaluated (i.e., 64, 128, 256, and \SI{1024}{\micro\meter}), beginning at either the outer edge of the expert tumor annotation (left column) or \SI{256}{\micro\meter} outside of the annotation (right column). Two foundation models were evaluated: UNI (top row) and Phikon (bottom row). Multiple halos of non-tumor tissue were sufficient to drive false positive metastasis detection. In the UNI-based model, for example, a \SI{64}{\micro\meter} halo beginning at the tumor annotation border from 7 positive specimens was sufficient to drive false positives. In the Phikon-based model, a \SI{1024}{\micro\meter} halo was sufficient for false positive predictions from 5 positive specimens.}
    \label{fig:supp_peritumoral}
\end{suppfigure}

\clearpage
\begin{suppfigure}[htbp]
\renewcommand\figurename{Supplementary~Figure}
\centering
\includegraphics[width=7in]{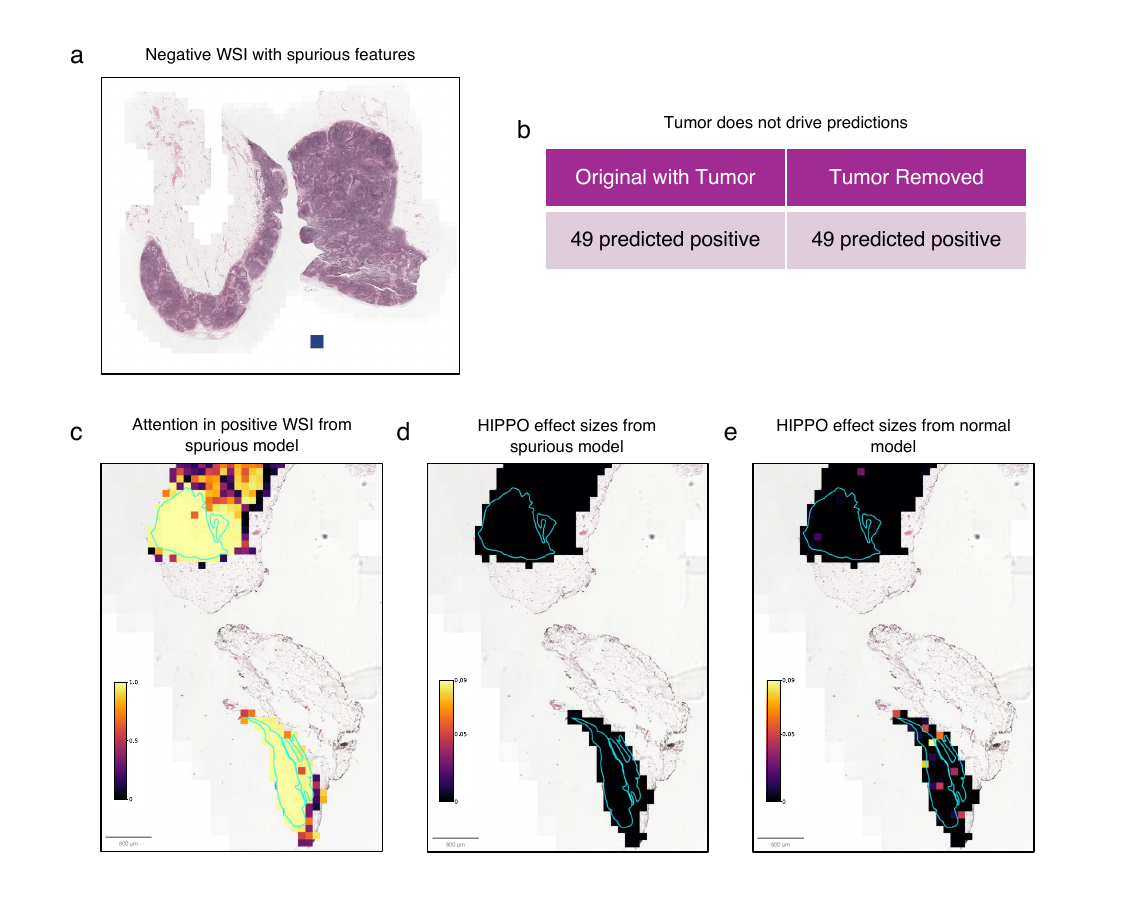}
\caption{\small{\textbf{HIPPO identifies shortcut learning when attention struggles.} \textbf{a,} Thumbnail of a negative specimen (\texttt{normal\textunderscore009}) with a $768 \times 768$ \SI{}{\micro\meter} blue square added. A blue square was added to all negatives specimens (n=239) in the CAMELYON16 dataset to promote shortcut learning. The UNI foundation model was used to embed the tissue and the blue squares. Positive samples were unaltered. 
\textbf{b,} All positive specimens were predicted as positive, and removal of tumor regions did not change model predictions. This suggested that the ABMIL models learned that if a blue patch is absent, the specimen is positive for metastasis.
\textbf{c,} Attention heatmap for specimen \texttt{test\textunderscore002}, with expert tumor annotation in cyan. Despite tumor having no effect on model predictions, there was strong attention on tumor regions.
\textbf{d,} Heatmap of patch effect sizes in specimen \texttt{test\textunderscore002} using the ABMIL model trained on deliberate spurious specimens. Using \textit{HIPPO-search-positive-effect}, we searched for the patches with the highest effect on model outputs.
\textbf{e,} Heatmap of patch effect sizes in specimen \texttt{test\textunderscore002} using the original ABMIL model, trained without deliberate spurious specimens.
}}
\label{fig:cam16-shortcuts}
\end{suppfigure}

\clearpage
\begin{suppfigure}[htbp]
\renewcommand\figurename{Supplementary~Figure}
    \centering
    \includegraphics[width=1.0\textwidth]{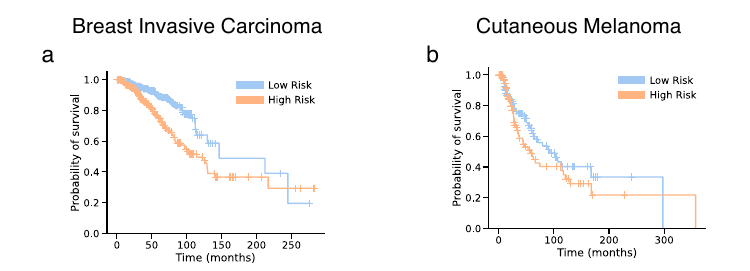}
    \caption{\textbf{a, b,} Kaplan Meier plots for breast cancer (BRCA) (\textbf{a}) and cutaneous melanoma (SKCM) (\textbf{b}) in The Cancer Genome Atlas. Prognostic attention-based multiple instance learning models were trained to learn overall survival from whole slide images (WSIs), and risk scores were used to stratify patients. If a patient had multiple WSIs, the predicted prognoses were averaged across WSIs to arrive at a single predicted risk score per patient. Risk scores were then median split into low risk and high risk. BRCA overall survival had concordance index of 0.667 ($p<0.005$, log-rank test), and for SKCM, concordance index was 0.557 ($p>0.05$, log-rank test). Please note that for experiments in the main text, low and high risk were defined as the first and fourth quartiles of risk scores, respectively.}
    \label{fig:supp_surv_km}
\end{suppfigure}

\clearpage
\begin{suppfigure}[htbp]
\renewcommand\figurename{Supplementary~Figure}
\centering
\includegraphics[width=\linewidth]{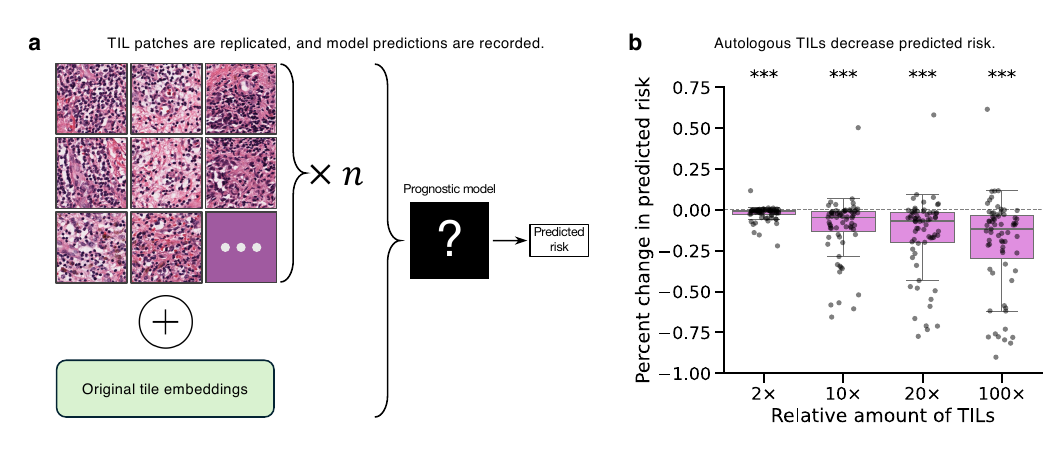}
\caption{\textbf{Autologous TILs improve predicted prognosis.} 
In high-risk slides of cutaneous melanoma (TCGA-SKCM, n=67), TIL-positive patches were identified using a heuristic from \cite{chen2022porpoise}. High risk was defined as slides with the top 25\% of predicted risk scores. 
\textbf{a,} The embeddings of TIL-positive regions were replicated and concatenated with the original embeddings (the ellipsis denotes that the displayed TIL patches are a representative sample of a larger set). Model predictions are then recorded for this counterfactual with additional autologous TILs.
\textbf{b,} Box plot showing the difference in model predictions, relative to the original specimens. Differences are shown on the $y$-axis and were calculated as the predicted risks with autologous TILs minus the original predicted risk (negative values indicate that autologous TILs decreased predicted risk). The $x$-axis shows the amount of TILs relative to the original specimens. The sample size in each box is 67.
Box plots show the first and third quartiles, the median (central line) and the range of data with outliers removed (whiskers), and significance is shown (***: $p < 0.001$).
}
\label{fig:supp_surv_autologous_tils}
\end{suppfigure}

\clearpage
\begin{suppfigure}[htbp]
\renewcommand\figurename{Supplementary~Figure}
    \centering
    \includegraphics[width=\linewidth]{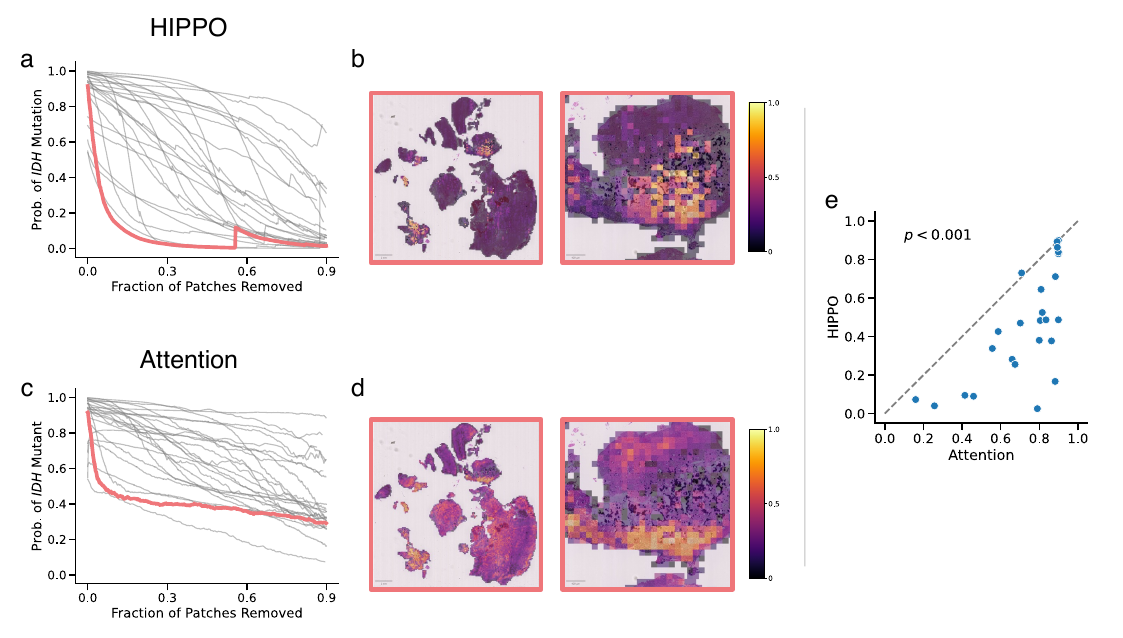}
    \caption{\textbf{HIPPO outperforms attention in identifying regions that drive positive predictions.} The strategy \texttt{HIPPO-search-positive-effect} was used to identify the patches that were most responsible for predictions for samples predicted as positive. Ten patches were removed at each iteration of the HIPPO search to reduce running time. For attention, ten patches were removed at a time, in order of descending attention, for comparison with HIPPO. \textbf{(a, c)}, line plots showing the probability of \textit{IDH} mutation on the vertical axis and the ratio of patches removed on the horizontal axis, where patches are removed by \textbf{(a)} HIPPO search  or \textbf{(c)} attention. \textbf{(b, d)}, heatmap of patches found by HIPPO \textbf{(b)} and heatmap of attention weights \textbf{(d)}, both normalized to range [0, 1]. \textbf{(e)} scatter plot showing the ratio of patches removed to decrease the predicted \textit{IDH} mutation to 0.4. HIPPO more effectively identified the patches driving positive predictions, requiring fewer patch removals to reduce the probability to 0.4 compared to attention ($p<0.001$, independent t-test).}
    \label{fig:supp_ebrains_pos}
\end{suppfigure}

\clearpage
\begin{suppfigure}[htbp]
\renewcommand\figurename{Supplementary~Figure}
    \centering
    \includegraphics[width=0.5\linewidth]{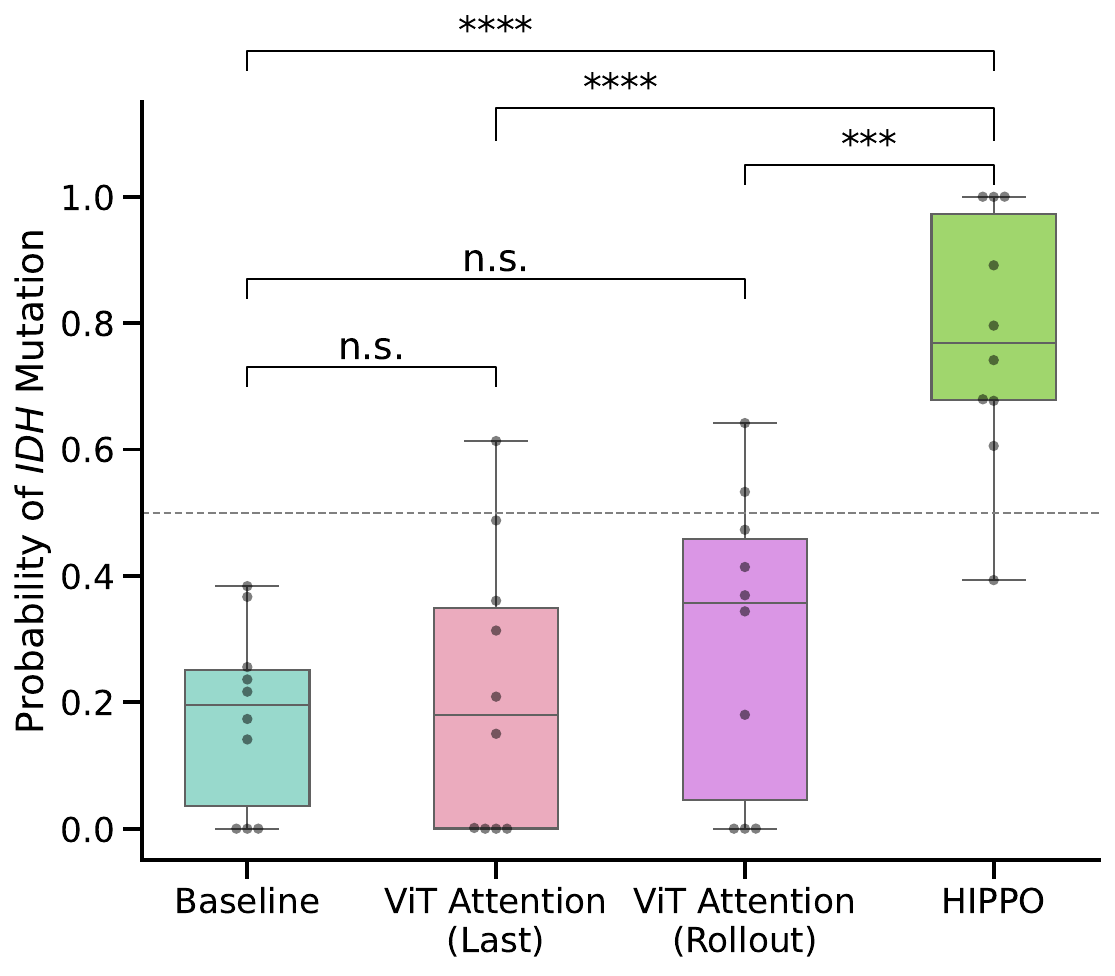}
    \\\rebuttaltemp{Added figure: Applied HIPPO to ViT and compared its performance to ViT-derived attention scores}
    \caption{\rebuttal{Box plot of model probabilities for \textit{IDH} mutation in FN specimens
at baseline (original specimens), ViT-based attention, and HIPPO. Removing the top 20\% patches found by HIPPO rescued true positive predictions in many
cases and led to significantly higher model probabilities than baseline ($p < 0.0001$, independent two-sided t-test) and both types of ViT-based attention ($p < 0.0001$,
independent two-sided t-test). Removing the top 20\% of patches by both ViT Attention (Last) and ViT Attention (Rollout) did not significantly change model predictions from baseline
($p > 0.05$, independent two-sided t-test). Classification threshold is shown at $y = 0.5$.}}
    \label{fig:supp_idh_prediction_with_vit}
\end{suppfigure}

\clearpage
\begin{suppfigure}[htbp]
\renewcommand\figurename{Supplementary~Figure}
    \centering
    \includegraphics[width=4in]{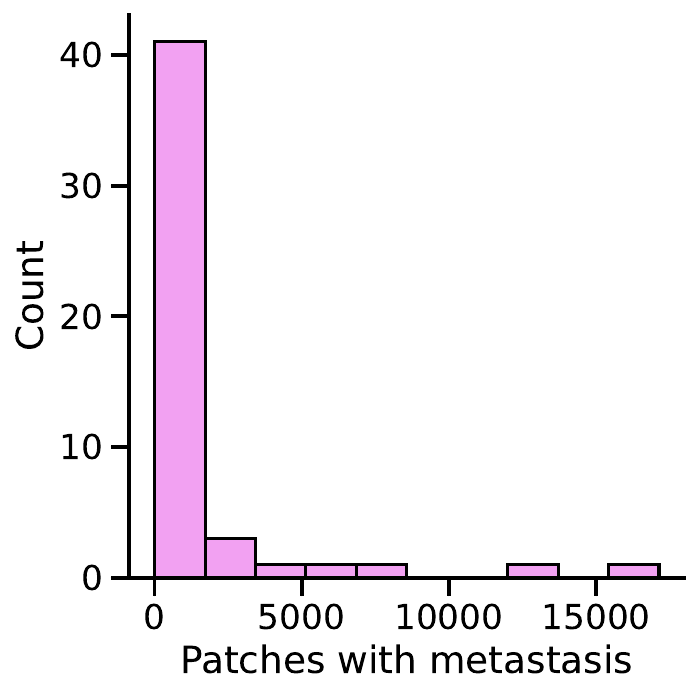}
    \caption{Distribution of the number of metastasis-containing patches in the CAMELYON16 test set. The test consists of 49 specimens with metastasis (plotted here) and 80 specimens negative for tumor. The positive specimens contained an average of 1320 patches that intersected the tumor annotation, where each patch was $128 \times 128$ \SI{}{\micro\meter}. This distribution was heavily right-skewed. The median number of tumor patches was 102, and the 25th and 75th percentiles were 18 and 616, respectively.}
    \label{fig:supp_hist_patches_with_mets}
\end{suppfigure}

\newpage

\phantomsection
\section*{Supplementary Methods}
\setcurrentname{Supplementary Methods} 
\label{sec:supplementary_methods}

\subsection*{HIPPO details}

\subsubsection*{Core mechanism of HIPPO}

HIPPO is an explainable AI toolkit designed to construct counterfactual whole slide images (WSIs) that are valid for multiple instance learning (MIL) models by manipulating patch embeddings rather than synthesizing new images. The key insight enabling HIPPO is that MIL models represent WSIs as bags of patches that are permutation-invariant—the model’s output remains unchanged regardless of the order of the patches. This property is crucial for HIPPO's functionality, as it allows for the manipulation of patches (addition or removal) without introducing unrealistic artifacts, such as gray pixel masking, that could confound model predictions.

In other words, within a MIL framework, a WSI is represented as a set of patch embeddings:
\[
P = \{p_1, p_2, \dots, p_N\},
\]
where each \(p_i \in \mathbb{R}^d\) is a feature vector extracted from a corresponding image patch using a pre-trained embedding model.

The MIL model aggregates these embeddings to produce a specimen-level prediction:
\[
f(P) = f(\{p_1, p_2, \dots, p_N\}) \in \mathbb{R},
\]
which is typically a continuous probability that can be thresholded to obtain a binary label.

Because MIL is permutation-invariant, its output does not depend on the ordering of the patches:
\[
f(P) = f(\{p_1, p_2, \dots, p_N\}) = f(\{p_{\sigma(1)}, p_{\sigma(2)}, \dots, p_{\sigma(N)}\}),
\]
for any permutation \(\sigma\). This permutation-invariance property is a fundamental characteristic of MIL models, as they treat WSIs as ``bags of patches'' where the order of patches does not influence the final prediction. This allows HIPPO to construct valid counterfactual WSIs by simply adding or removing patches from the bag, without performing image synthesis or inpainting. These resampled bags remain valid inputs—the model naturally accommodates any combination of patches (Fig. \ref{fig:hippo-schematic} and Supplementary Fig. \ref{fig:hippo-schematic-insert-delete}). 

HIPPO quantifies the contribution of specific patches to the overall prediction by comparing model predictions before and after modification:
\[
f(P \setminus P_{\text{removed}}) - f(P),
\quad \text{or} \quad
f(P \cup P_{\text{added}}) - f(P)
\]
where $P_{\text{removed}}$ denotes the subset of patch embeddings that are excluded from the original bag $P$ to simulate the removal of certain regions (Supplementary Fig. \ref{fig:hippo-schematic-insert-delete}a), and $P_{\text{added}}$ denotes the set of additional patch embeddings that are introduced into the original bag $P$ to simulate the inclusion of new histological evidence (Supplementary Fig. \ref{fig:hippo-schematic-insert-delete}b).

This mechanism circumvents the challenges of direct image manipulation while exploiting the structural properties of multiple instance learning to generate interpretable and quantitative insights into model behavior.

\subsubsection*{HIPPO usage}

HIPPO provides four methods for intervention selection: \textit{HIPPO-knowledge}, which is based on a user-defined hypothesis, \textit{HIPPO-attention}, and two data-driven approaches, such as \textit{HIPPO-search-positive-effect}, and \textit{HIPPO-search-negative-effect}. All methods share the same underlying principle---measuring how model predictions change when specific patches are added to or removed from a WSI. However, they differ in how the patches to add or remove ($P_{\text{added}}$ or $P_{\text{removed}}$) are chosen and whether the removal (or addition) is performed in a single step or sequentially according to user-defined criteria.

\subsubsection*{HIPPO-knowledge}
\textit{HIPPO-knowledge} quantifies the effect of regions identified based on prior knowledge
(e.g., expert annotations).
Patches corresponding to annotated regions are removed or added
to test their necessity or sufficiency for model predictions.

\begin{algorithm}[H]
\caption{HIPPO-knowledge}
\begin{algorithmic}[1]
\Statex
\textbf{Input:}
\begin{itemize}
    \item $P_{\text{target}}$: Target set of patch embeddings. We perform an intervention on this set. 
    \item $P_\text{source}$: Source set of patch embeddings for which we have expert annotations. For ``remove'' mode, we assume $P_{\text{source}} \subset P_{\text{target}}$.
    \item $f(\cdot)$: ABMIL model mapping patch embeddings to slide-level prediction
    \item $A(\cdot)$: Annotation function returning indices of the region of interest
    \item $\text{type}$: Intervention type, either ``remove'' or ``add''
\end{itemize}
\Statex
\textbf{Output:}
\begin{itemize}
    \item $s_{\text{baseline}}$: Model prediction before intervention
    \item $s_{\text{modified}}$: Model prediction after intervention
    \item $\text{effect}$: Difference between baseline and modified predictions
\end{itemize}
\Statex
\hrulefill
\If{$\text{type} = \text{``remove''}$} 
    \State $s_{\text{baseline}} \gets f(P_\text{target})$ \Comment{Get original prediction with all patches}
    \State $P_{\text{annotated}} \gets \{p_i : i \in A(P_\text{source})\}$ \Comment{Extract patches in annotated region}
    \State $P_{\text{modified}} \gets P_\text{target} \cap  P_{\text{annotated}}^c$ \Comment{Remove annotated patches from target slide}
    \State $s_{\text{modified}} \gets f(P_{\text{modified}})$ \Comment{Get prediction after intervention}
    \State $\text{effect} \gets s_{\text{baseline}} - s_{\text{modified}} $     
\ElsIf{$\text{type} = \text{``add''}$} 
    \State $s_{\text{baseline}} \gets f(P_{\text{target}})$ \Comment{Get baseline prediction from target slide}
    \State $P_{\text{annotated}} \gets \{p_i : i \in A(P_\text{source})\}$ \Comment{Extract patches from annotated region of source}
    \State $P_{\text{modified}} \gets P_\text{target} \cup P_{\text{annotated}}$ \Comment{Add annotated patches to target slide}
    \State $s_{\text{modified}} \gets f(P_{\text{modified}})$ \Comment{Get prediction after intervention}
    \State $\text{effect} \gets s_{\text{modified}} - s_{\text{baseline}}$     
\EndIf
\State \Return $(s_{\text{baseline}}, s_{\text{modified}}, \text{effect})$
\end{algorithmic}
\end{algorithm}

\subsubsection*{HIPPO-attention}
\textit{HIPPO-attention} quantifies the effect of high-attention regions
by removing top-attention patches and measuring the resulting change in model prediction.
This provides both the direction and magnitude of influence for attention-based models.

\begin{algorithm}[H]
\caption{HIPPO-attention}
\begin{algorithmic}[1]
\Statex
\textbf{Input:}
\begin{itemize}
    \item $P = \{p_1, p_2, \dots, p_N\}$: A set of patch embeddings from a WSI
    \item $f(\cdot)$: MIL model mapping patch embeddings to slide-level prediction
    \item $\text{Attention}(\cdot)$: Function returning attention scores derived from the MIL model $\alpha = \{\alpha_1, \dots, \alpha_N\}$
    \item $k$ or $f$: Threshold for selecting top patches (e.g., top $k$ or fraction $f = 0.01$)
\end{itemize}
\Statex
\textbf{Output:}
\begin{itemize}
    \item $s_{\text{baseline}}$: Model prediction before intervention
    \item $s_{\text{modified}}$: Model prediction after removing high-attention patches
    \item $\text{effect}$: Difference between baseline and modified predictions
\end{itemize}
\Statex
\hrulefill
\State $s_{\text{baseline}} \gets f(P)$ \Comment{Get original prediction with all patches}
\State $\alpha \gets \text{Attention}(P)$ \Comment{Extract attention scores for each patch}
\State $I_{\text{high}} \gets \text{TopK}(\alpha, k)$ or $\text{TopFraction}(\alpha, f)$ \Comment{Identify indices of highest-attention patches}
\State $P_{\text{high}} \gets \{p_i : i \in I_{\text{high}}\}$ \Comment{Collect the high-attention patches}
\State $P_{\text{modified}} \gets P \cap P_{\text{high}}^c$ \Comment{Remove high-attention patches from slide}
\State $s_{\text{modified}} \gets f(P_{\text{modified}})$ \Comment{Get prediction without high-attention patches}
\State $\text{effect} \gets s_{\text{baseline}} - s_{\text{modified}}$ 
\State \Return $(s_{\text{baseline}}, s_{\text{modified}}, \text{effect})$
\end{algorithmic}
\end{algorithm}

\subsubsection*{HIPPO greedy patch removal search}
Unlike HIPPO-knowledge, which requires a prior assumption on which subregion would be influential, HIPPO greedy search is a data-driven approach that identifies high-impact patches through systematic exploration. The algorithm iteratively removes patches and measures the resulting change: it temporarily removes each remaining patch, measures how the prediction changes, then permanently removes the patch that best matches the search objective. This process repeats iteratively, building an ordered sequence that reveals the relative importance of different patches. Two search objectives are available depending on the research question:
\begin{itemize}
    \item HIPPO-search-positive-effect: a greedy search algorithm to identify the regions that most strongly drive a prediction to the positive direction. We apply this method to samples predicted as positive by the model to identify regions influential for their predictions (Fig. \ref{fig:hippo-schematic}e).
    \item HIPPO-search-negative-effect: a greedy search algorithm to identify the regions that most strongly drive a prediction to the negative direction. We apply this method to samples predicted as negative by the model to identify regions influential for their predictions.
\end{itemize}

We apply HIPPO-search-positive-effect to samples predicted as positive by the model to identify regions influential for their predictions (Fig. \ref{fig:hippo-schematic}e). We apply HIPPO-search-negative-effect to samples predicted as negative by the model to identify regions influential for their predictions.

\begin{algorithm}[H]
\caption{HIPPO greedy patch removal search}
\begin{algorithmic}[1]
\Statex\textbf{Input:}
\Statex\begin{itemize}
    \item $P = \{p_1, p_2, \dots, p_N\}$: A set of patch embeddings from a WSI
    \item $f(\cdot)$: MIL model mapping patch embeddings to slide-level prediction
    \item $\pi$: Patch selection function ($\pi_{\text{positive}}$ or $\pi_{\text{negative}}$)
    \begin{itemize}
    \item For HIPPO-search-positive-effect, we use $\displaystyle \pi_{\text{positive}}(\{(i,s_i)\}) = \arg\max_{i} s_i $  
    \item For HIPPO-search-negative-effect, we use $\displaystyle \pi_{\text{negative}}(\{(i,s_i)\}) = \arg\min_{i} s_i $ 
    \end{itemize}
    \item $R$: Stopping criterion (e.g., number of iterations or fraction of patches removed)
\end{itemize}
\Statex
\textbf{Output:}
\begin{itemize}
    \item $A$: Ordered list of removed patches
    \item $O$: Ordered list of model outputs after each removal
    \item $S$: Ordered list of effect sizes of removed patches
\end{itemize}
\Statex
\hrulefill
\State $A \gets [\,]$ \Comment{Initialize empty removal sequence}
\State $s^{(0)} \gets f(P)$ \Comment{Baseline prediction with all patches}
\State $O \gets [o^{(0)}]$ \Comment{Initialize prediction history list}
\State $S \gets []$ \Comment{Initialize effect size list}
\State $I \gets \{1, 2, \dots, N\}$ \Comment{Set of remaining patch indices}
\For{$r = 1$ to $R$} \Comment{Iterate until stopping criterion reached}
    \If{$|I| \le 1$} \Comment{Stop if only one or no patches remain}
        \State \textbf{break}
    \EndIf
    \For{each $i \in I$} \Comment{Evaluate effect of removing each remaining patch}
        \State $s_i \gets f(P) - f(P \cap \{p_i\}^c)$ \Comment{Predict with patch $i$ removed and compute the effect size}
    \EndFor
    \State $i^\star \gets \mathcal{S}(\{(i, s_i) : i \in I\})$ \Comment{Select patch to remove using optimizer}
    \State $A \gets A \cup [i^\star]$ \Comment{Add selected patch to removal sequence}
    \State $I \gets I \cap \{i^\star\}^c$ \Comment{Remove patch index from remaining set}
    \State $P \gets P \cap \{p_{i^\star}\}^c$ \Comment{Permanently remove patch from slide}
    \State $S \gets S \cup \{s_{i^\star}\}$ \Comment{Permanently remove patch from slide}
    \State $O \gets O \cup [o_{i^\star}]$ \Comment{Record prediction after removal}
\EndFor
\State \Return $(A, O, S)$ 
\end{algorithmic}
\end{algorithm}




\printbibliography[]
\end{refsection}

\end{document}